\pdfoutput=1
\documentclass[%
 reprint,
 amsmath,amssymb,
 aps,
]{revtex4-1}

\usepackage{graphicx}
\usepackage{dcolumn}
\usepackage{bm}

\usepackage[caption=false]{subfig}
\usepackage{multirow}

\begin{document}
\title{Density shocks in the relativistic expansion of highly charged one component plasmas}
\author{B. Zerbe}\email{zerbe@msu.edu}
\author{P. M. Duxbury}\email{duxbury@msu.edu}

\affiliation{Department of Physics and Astronomy, Michigan State University}
\date{\today}

\begin{abstract}
In a previous paper we showed that dynamical density shocks  occur in 
the non-relativistic expansion of dense single component plasmas relevant to ultrafast electron 
microscopy; and we showed 
that fluid models capture these effects accurately.  
We show that the non-relativistic decoupling of the relative and center of mass motions 
ceases to apply and this coupling leads to novel behavior in the relativistic dynamics
under planar, cylindrical, and spherical symmetries.   In cases 
where the relative motion of the bunch is relativistic, we show that a dynamical shock emerges
even in the case of a uniform bunch with cold initial conditions;  and that 
density shocks are in general enhanced when the relative motion becomes relativistic. 
Furthermore, we examine the effect of an extraction field on the relativistic dynamics
of a planar symmetric bunch.  
\end{abstract}
\maketitle

\section{Introduction}

 The expansion dynamics of highly charged plasmas is a fundamental problem 
in areas ranging from astrophysics to nanotechnology to beam physics.  
Previous analytic work 
has focused on initial conditions where a highly charged plasma is cold and has uniform 
density\cite{Jansen:1988_book,Reiser:1994_book,Batygin:2001_self,Bychenkov:2005_coulomb_explosion,Grech:2011_coulomb_explosion,Kaplan:2003_shock,Kovalev:2005_kinetic_spherically_coulomb_explosion,Last:1997_analytic_coulomb_explosion,Eloy:2001_coulomb_explosion,Krainov:2001_ce_dynamics,Morrison:2015_slow_down_dynamics,Boella:2016_multiple_species,Bychenkov:2011_relativistic}
However, the vast majority of this work, with the exception of Bynchenkov and 
Kovalev\cite{Bychenkov:2011_relativistic}, have assumed non-relativistic conditions.
In ultrafast electron microscopy (UEM)
and some beam physics applications, electron sources 
are used to produce dense bunches of charged particles within
an intense extraction field that is used to
accelerate the distribution to near-luminal speeds.  
In addition, for sufficient densities, the bunch self-field can result in relativistic
velocities within the frame of the bunch.  These concerns indicate that a relativistic theory is 
required in many practical cases, and here we present the relevant theory.

The typical analytic approach to relativistic expansion dynamics of such systems is a
treatment based on envelope equations that are predicated on the conservation of emittance and the use of 
uniform density distributions\cite{Reiser:1994_book}.  Uniform 
ellipsoidal distributions are particularly amenable to analysis as the self-electrostatic field 
in these distributions is linear and the
expansion dynamics results in a simple power law growth of the ellipsoid  axes -- at least
in the non-relativistic regime.  Furthermore, it is fairly straightforward to show that uniform ellipsoids
conserve emittance as long as all the particles can be treated as having identical 
Lorentz factors.
Moreover, analysis of beam dynamics such as emittance 
oscillation\cite{Anderson:1987_emittance}, 
emittance compensation\cite{Rosenzweig:2006_emittance_compensation}, 
and the beam halo \cite{Gluckstern:1994_analytic_halo}
generally assume similar
uniform-like 
conditions.  However, for electron injectors utilizing photoemission the initial conditions of the bunch is 
often Gaussian, or at the very least non-uniform, and it has long been known that 
charge redistribution from the non-uniform to the uniform bunch is one of the major sources of emittance 
growth\cite{Wangler:1991_emittance_growth} suggesting that uniform
distributions are at best an idealization that miss much of the physics
present in the typical situation.  However, we have recently shown that 
dynamics similar to Wangler's charge redistribution for freely expanding bunches leads to an opportunity of
``Coulomb cooling" --- a mechanism we believe employs the intense Coulomb fields
to carry off heat from non-neutral 
plasmas\cite{Williams:2017_transverse_emittance,Zerbe:2018_coulomb_dynamics}.
Our analysis is directed at
a better understanding of the relativistic expansion dynamics of uniform and non-uniform systems 
to study mechanisms of 
emittance growth near the particle source but also with the goal of understanding
and optimizing Coulomb cooling.  Here we concentrate on characterizing the relativistic density dynamics for 
both uniform and non-uniform initial conditions. 

Numerous works within the UEM literature have already looked at various aspects of the 
evolution of non-uniform distributions\cite{Degtyareva:1998_gaussian_pileup,Luiten:2004_uniform_ellipsoidal,Musucemi:2008_generate_uniform_ellipsoid,Morrison:2013_measurement,Li:2008_quasiellipsoidal,Siwick:2002_mean_field,Qian:2002_fluid_flow,Reed:2006_short_pulse_theory,Collin:2005_broadening,Gahlmann:2008_ultrashort,Tao:2012_space_charge,Portman:2013_computational_characterization,Portman:2014_image_charge,Michalik:2006_analytic_gaussian,Zerbe:2018_coulomb_dynamics}.  
Reed presented a fluid model that described the dynamics of non-uniform
bunch expansion under the assumptions that the bunch could be treated as having 
planar symmetry in the non-relativistic regime\cite{Reed:2006_short_pulse_theory}. 
In our previous paper\cite{Zerbe:2018_coulomb_dynamics}, 
we showed that the expansion dynamics in the planar case differs greatly from 
symmetries in higher dimensions and verified analytic description of the dynamics  
utilizing both N-particle and particle-in-cell (PIC) simulations.
The analytic descriptions
depend on terms that can be written as functions multiplied by the quantity  
\begin{align}
  D_{01} = \frac{{\bar{\rho}}_{01}}{\rho_{01}} - 1
\end{align} 
for planar geometries or by the quantity 
\begin{align}
  D_{0d} = \frac{\rho_{0d}}{{\bar{\rho}}_{0d}} - 1 \label{eq:D_{0d}}
\end{align}
for cylindrical ($d=2$) and spherical ($d=3$) geometries,
where $\rho_{0d}$ is the initial density at location $r_0$ and ${\bar{\rho}}_{0d}$ is the average
density within that location.  For the uniform distribution, the local density and the average density are the same so that $D_{0d} = 0$
everywhere and the density evolution's dependence on the aforementioned functions vanishes
 reducing the dynamics to the uniform dynamics utilized extensively in the literature.  
Otherwise, these functions play a large role in the density evolution leading to differences
in the dynamics of distributions with different initial density profiles.  Specifically, we showed that 
a density shock, seen in Coulomb explosion 
studies\cite{Grech:2011_coulomb_explosion,Kaplan:2003_shock,Kovalev:2005_kinetic_spherically_coulomb_explosion,Last:1997_analytic_coulomb_explosion,Murphy:2014_cold_ions,Reed:2006_short_pulse_theory,Degtyareva:1998_gaussian_pileup}, 
was present in the analytic density evolution of initially Gaussian distributions
under both cylindrical and spherical symmetries but absent under planar symmetry
unless an appropriate initial chirp in phase space is present.  Moreover, such
a shock is absent in non-relativistic dynamics of uniform systems with cold 
initial conditions suggesting that uniform distribution evolution is
unique in this regard\cite{Zerbe:2018_coulomb_dynamics}.   

We noted in our previous work that the density evolution seen in PIC simulations
using an electromagnetic (EM) solver and relativistic particle pusher, which should capture
all relativistic effects if the initial field is accurate, do not significantly
differ from our analytic description for the
densities analyzed there\cite{Zerbe:2018_coulomb_dynamics}.   
Moreover, we have seen that the PIC simulations
using an EM solver do not significantly differ from those using an 
electrostatic (ES) solver with relativistic particle
pusher for much higher densities than those examined in our previous work, or in the work described here.
  Assuming the accuracy of the 
initial fields, we conclude that the relativistic effects 
are then adequately captured within the relativistic particle pusher, which is equivalent
to simply including the relativistic momentum in the analysis.
Precisely such an analysis of the relativistic free-expansion of a 
spherically-symmetric, cold uniform
charge distribution was completed by Bychenkov and Kovalev\cite{Bychenkov:2011_relativistic},
and part of what we do in this manuscript is extend 
this analysis to non-uniform cases as well as additional symmetries.  

Here we treat charge distributions with general initial spatial distributions starting from rest 
under planar, cylindrical and 
spherical geometries introducing a novel length scale that is associated with each 
symmetry.   
First in Section \ref{subsec:gen}, we present general results applicable to all cases.
In Section \ref{subsec:planar} we derive expressions for planar symmetry for any arbitrary initial spatial distribution
and examine these expressions in the non-relativistic and highly relativistic limits.  
We then introduce $M$-shell simulations,
which are simulations of $M$ equally charged planes in 1D, and show that these
simulations reproduce the density evolution derived analytically.  In Sections
\ref{subsec:cyl} and \ref{subsec:sph}, we 
derive relativistic density evolution expressions under cylindrical and spherical symmetries,
respectively, 
for arbitrary initial distributions, and we show that these expressions are consistent with PIC 
calculations using an EM solver and relativistic particle pusher utilizing the well-known package 
warp\cite{Friedman:2014_warp}.  Further, we show that $M$-shell simulations, which 
track $M$ equally charged cylindrical- and spherical- shells in 2- and 3- dimensions,
respectively, also capture the same density evolution.  We validate
these expressions against their non-relativistic and uniform relativistic 
counterparts, and we examine the expressions in the highly relativistic limit.  

In Section \ref{sec:Ea} 
we introduce an external extraction field in the case of a planar electron bunch, and 
we point out that the acceleration from self-fields and external fields do not decouple in the 
relativistic case.   Though the analysis is captured by a straightforward extension of the analysis 
used for the planar case with no extraction field, the physical effects are quite interesting and relevant 
to ultrafast electron microscopy.  In Section \ref{sec:applied}, we demonstrate how to apply 
the time dependent distributions to calculate the statistical width of the distribution.  
Section \ref{sec:discussion} 
contains discussions and conclusions; noting in particular 
the emergence of density shocks due to relativistic effects, even in planar uniform systems where a shock 
does not emerge in the non-relativistic limit. 

\section{Relativistic density evolution}\label{sec:rest}

In this section we consider planar, cylindrical and spherical geometries 
in the case of cold initial conditions and where there is no external 
electromagnetic field.    We treat the dynamics using the 
relativistic treatment of momentum and energy, however we treat the 
forces between electrons using electrostatics.  We then check the latter approximation 
under cylindrical and spherical geometries using PIC calculations using a full EM solver and 
find excellent agreement. We start with a general analysis 
and then specialize in the later three subsections. 

\subsection{General considerations}\label{subsec:gen}

\subsubsection{General formulation}
We consider cold symmetric initial charge distributions so that the electric 
field at position $z_0$ in a planar geometry is given by,
\begin{equation}
  E_0(z_0) = E_{01} = \frac{\Sigma_{tot} P_{01}}{2 \epsilon_0} = E_{T1}  P_{01}
\end{equation}
where $\Sigma_{tot} = N q/ A$, with $A$ the pulse area, $N$ the number of particles 
in the bunch, $q$ the charge of each particle, 
$E_{T1} = \frac{\Sigma_{tot}}{2 \epsilon_0}$ is the total planar field produced by the particles,
and $P_{01}$ is a position specific
scalar representing the fraction of the electrons in the bunch contributing to the net electric field on the 
particle at position $z_0$.   The physical significance of the total field $E_{T1}$ will be discussed in 
detail in Section \ref{sec:Ea}.    The direction along which the charge density 
varies is $z$, and we assume the initial charge distribution is symmetric about the origin for the sake
of simplicity.    We define $\rho_1(z;t)$ to be
the number density per unit length at position $z$ and 
$\rho_{01} = \rho_{01}(z_0) =\rho_1(z_0;t=0)$ to be the initial 
number density at position $z_0$.   
As the initial density $\rho_{01}$ is symmetric about the origin, 
the cumulative probability is given by,
\begin{equation}
  P_{01} =   P_{01}(z_0) = \int_0^{z_0} 2 \rho_{01}(z) dz
\end{equation} 
Notice that the quantity $\Sigma_{tot} P_{01}$ 
represents the charge per unit area after integrating the number density per unit length $\rho_{01}$ over
the range $[-z_0,z_0]$.

In systems with cylindrical symmetry, we have, 
\begin{equation}
E_0(r_0) = E_{02} = { \Lambda_{tot} P_{02} \over 2\pi \epsilon_0 r_0} = E_{T2} P_{02}
\end{equation}
where $\Lambda_{tot}$ is the total charge per unit length along the 
cylindrical charge distribution, $E_{T2} = \frac{\Lambda_{tot}}{2\pi \epsilon_0 r_0}$
is the electric field a particle at $r_0$ would feel if the entire distribution were distributed cylindrically
within $r_0$, and $P_{02}$ is the cumulative probability given by,
\begin{equation}
P_{02} =  \int_0^{r_0} 2\pi r \rho_{02}(r) dr
\end{equation} 
where $\rho_2(r;t)$ is the number density in two dimensions (number per unit area) and 
where we define $\rho_{02} = \rho_{02}(r_0) = \rho_2(r_0;t=0)$.
Notice that the quantity $\Lambda_0 P_{02} $ represents the
charge per unit length inside radius $r_0$. 

In systems with spherical symmetry we have 
\begin{equation}
E_0(r_0) = E_{03} = {Q_{tot} P_{03}\over 4\pi \epsilon_0 r_0^2} = E_{T3} P_{03}
\end{equation}
where $Q_{tot}$ is the total charge in the system, 
$E_{T3} = \frac{Q_{tot}}{4 \pi \epsilon_0 r_0^2}$ and is the electric field a particle at $r_0$ 
would feel if the entire distribution were distributed spherically within
$r_0$, and $P_{03}$ is the cumulative probability given by,
\begin{equation}
P_{03} =  \int_0^{r_0} 4\pi r^2 \rho_{03}(r) dr
\end{equation}
where $\rho_3(r;t)$ is the number density in three dimensions (number per unit volume) and 
where we define $\rho_{03} = \rho_{03}(r_0) = \rho_3(r_0;t=0)$.
Again notice that 
$P_{03}$ represents the fraction of the particles that lie inside radius $r_0$ and $Q_{tot} P_{03}$ gives 
the charge inside radius $r_0$. 

To make analytic progress we make the laminar fluid approximation, which 
states that there is no mixing of the charged particle trajectories.  As a result, the symmetries 
of the charge distributions are conserved.   If we consider a particle of charge $q$ and 
rest mass $m$ starting from rest (cold initial conditions); at position $z_0$ (planar case) 
or $r_0$ (cylindrical or spherical cases),  we want to find the position and velocity of the 
particle at later times.   In a planar system we thus 
want to find $z(z_0,t)=z$ and $v(z_0,t) = v$; while in the cylindrical and spherical cases we want to find 
the radial position and radial velocity $r(r_0,t) = r$; $v(r_0,t) = v$.   One approach is to 
use the relativistic form of Newton's second law, 
\begin{equation}
  \frac{dp}{dt} = q E
\end{equation}
where we are considering the $z$ component of these vectors in planar geometries 
and the radial component in cylindrical and spherical geometries.   We use the relativistic 
expression for the momentum, 
\begin{equation}
  p = \gamma m v
\end{equation}
where $m$ is the rest mass and 
\begin{equation}
  \gamma = \frac{1}{\sqrt{1-(\frac{v}{c})^2}}
\end{equation}
Due to the laminar fluid property, the electric field experienced by a particle at position $z(z_0,t)$ 
in planar geometries is the same as the electric field this particle experienced at it's 
initial position, so that
\begin{equation}
  E_1(z) = E_{01}
\end{equation}
In cylindrical geometries using Gauss' law we find,
\begin{equation}
  E_2(r) = E_{02} \frac{r_0}{r}
\end{equation}
and analogously in spherical geometries we have, 
\begin{equation}
  E_3(r) = E_{03} \left(\frac{r_0}{r}\right)^2
\end{equation}
These expressions may be used with Newton's second law to solve for the 
particle dynamics.  Alternatively in an energy formulation, conservation of 
energy requires that the change in kinetic energy equals the change in 
potential energy.   We use the relativistic form of the kinetic energy,
\begin{equation}
  K = (\gamma -1) mc^2
\end{equation}
and the change in kinetic 
energy is equal to $K$ as 
the bunch starts from rest.   The change in potential energy is found 
by integrating the force $qE$, which for the planar case gives, 
\begin{equation}
  \Delta U_1 = q E_{01} (z_0-z) \label{eq:U1}
\end{equation}
while for the cylindrical case we find,
\begin{equation}
  \Delta U_2 = q E_{02}r_0 \ln\left(\frac{r_0}{r}\right)  \label{eq:U2}
\end{equation}
and for the spherical case,
\begin{equation}
  \Delta U_3 = q E_{03} r_0^2 \left(\frac{1}{r} - \frac{1}{r_0}\right)  \label{eq:U3}
\end{equation}
Notice that $\Delta U \le 0$ for all time due to the physics as the potential 
energy decreases as the electron bunch expands.
Setting the sum of the changes in potential and kinetic energy to 
zero, we have, 
\begin{equation}
  (\gamma-1) mc^2 = - \Delta U\label{eq:KE gen}
\end{equation}
where the appropriate expression for $\Delta U$ must be utilized.  From 
this expression a general relation between the velocity and position is 
found to be, 
\begin{equation}
  v = \frac{\sqrt{ \left(1- {\Delta U\over mc^2}\right)^2 - 1}}{1 - \frac{\Delta U}{mc^2}} c\label{eq:v}.
\end{equation}
Moreover, since $\Delta U$ only depends on the position, this equation may be 
integrated to find an expression relating time and position,
\begin{equation}
  t-t_0 = \int_{y_0}^y \frac{1}{v} dy\label{eq:t}
\end{equation}
where $y=z$ for planar cases and $y=r$ for cylindrical and spherical cases.  

Finally to obtain an expression for the time evolution of the density, we 
use the conservation of the charge density under laminar conditions. This conservation can be stated 
for planar systems as
\begin{equation}
  \rho_{01} dz_0 = \rho_1(z;t) dz
\end{equation}
for cylindrical geometries as
\begin{equation}
  \rho_{02} ~ 2 \pi r_0 dr_0 = \rho_2(r;t) ~2 \pi r dr
\end{equation}
and for spherical systems as
\begin{equation}
  \rho_{03} ~ 4 \pi r_0^2 dr_0 = \rho_3(r;t) ~4 \pi r^2  dr
\end{equation}
In general, this results in the relationship between the density and the initial density of
\begin{align}
  \rho_d(y;t) &=  \frac{\rho_{0d}}{\left(\frac{y}{y_0}\right)^{d-1}y'},\label{eq:den gen}
\end{align}
where again $y=z$ for planar cases and $y=r$ for cylindrical and spherical cases, $d$ is
$1$, $2$, and $3$ for these symmetries, respectively, and $' \equiv \frac{d}{dy_0}$ with
the $d$'s in this last expression representing differentiation -- not dimensionality of the problem.

\subsubsection{Fundamental parameters}
Some fundamental parameters need to be considered in the discussion of 
high density single component plasmas.  The first is the plasma frequency, 
\begin{equation}
  \omega_p = \sqrt{{q^2 n \over \epsilon_0 m}}\label{eq:rel wp}
\end{equation}
which describes the frequency of coherent plasma oscillations, where $n$ is the 
number density (number of particles per unit volume) and $q$ the particle charge.  We note that relativistic
effects affect the plasma frequency, but as our distribution is starting from rest, it is sufficient
to consider the non-relativistic plasma frequency; however, the plasma frequency for different
symmetries is not apparent from Eq. (\ref{eq:rel wp}). 
We define average initial densities $\overline{\rho}_{01} = \frac{P_{01}}{2 z_0}$,
 $\overline{\rho}_{02} = \frac{P_{02}}{\pi r_0^2}$,  $\overline{\rho}_{03} = \frac{P_{03}}{\frac{4}{3} \pi r_0^3}$
which are the average densities inside distance $z_0$ (planar case), or inside radius $r_0$ for the 
cylindrical and spherical cases.  These definitions are used to define initial plasma frequencies as 
follows for planar systems, 
\begin{equation}
  \omega_{01} = \sqrt{{q \Sigma_{tot} \overline{\rho}_{01}  \over \epsilon_0 m }} = \sqrt{\frac{q E_{01}}{m z_0}}\label{eq:omega01},
\end{equation}
cylindrical systems,
\begin{equation}
  \omega_{02} = \sqrt{\frac{q \Lambda_{tot} \overline{\rho}_{02}}{\epsilon_0 m }} = \sqrt{\frac{2 q E_{02}}{m r_0}}\label{eq:omega02},
\end{equation}
and spherical systems
\begin{equation}
  \omega_{03} = \sqrt{\frac{q Q_{tot} {\bar{\rho}}_{03}}{\epsilon_0 m }} = \sqrt{\frac{3 q E_{03}}{m r_0}}\label{eq:omega03}.
\end{equation}
This can be summarized by
\begin{align}
  \frac{\omega_{0d}^2}{c^2} = \frac{d}{y} \frac{q E_{0d}}{m c^2}\label{eq:omega_d}
\end{align}
for $d \in \{1,2,3\}$ and where $y$ is $z_0$ when $d = 1$ and $r_0$ for $d = 2$ and $d=3$.

As will be seen below, the time $\tau_{0d}$ defined as,
\begin{equation} 
\tau_{0d} = \frac{2\pi}{\omega_{0d}}
\end{equation}
 sets the timescale 
for the relativistic expansion of high density charge clouds; as was found 
 in the non-relativistic cases\cite{Zerbe:2018_coulomb_dynamics}.  

In addition to the plasma frequency, we find it advantageous to define the related 1D-number density as 
\begin{align}
  \rho_{r0d} &=  \frac{q E_{0d}}{mc^2} = \frac{q E_{Td} P_{0d}}{mc^2} \label{eq:density scale}
\end{align}
where  $\rho_{r0d}$ has units of inverse length. 
We call $\rho_{r0d}$ the relativistic crossover density
for planar, cylindrical, and spherical symmetries for $d = 1,2,3$, respectively.
The physical interpretation of the relativistic crossover density 
is that it provides a scale for the potential energy as 
$\frac{\Delta U_{1}}{mc^2} = \rho_{r01}(z_0-z)$,
$\frac{\Delta U_{2}}{mc^2} =  \rho_{r02}r_0\ln\left(\frac{r_0}{r}\right)$, and
$\frac{\Delta U_{3}}{mc^2} =  \rho_{r03}r_0\left(\frac{r_0}{r} - 1\right)$.
The relativistic crossover density is related to  the plasma frequency through
\begin{align}
  \rho_{r0d} &=  \frac{y}{d}\frac{\omega_{0d}^2}{c^2}\label{eq:density scale}.
\end{align}
 The relativistic length scale, $l_{r0d}$ is related to the relativistic density through,
\begin{align}
  l_{r0d} &=  \frac{P_{0d}}{\rho_{r0d}}\nonumber\\
              &= \frac{mc^2}{q E_{Td}}\label{eq:length scale}
\end{align}
where $l_{r0d}$ is seen to be independent of the initial distribution.  
$l_{r0d}$ can be thought of as the distance a particle experiencing the force obtained by
the full distribution at the given coordinate 
needs to travel before having kinetic 
energy of $m c^2$.  Notice, that $l_{r01}$ is
a constant and is specifically independent of $z_0$; however, $l_{r02} \propto r_0$ and
$l_{r01} \propto r_0^2$. 

\subsection{Planar symmetry}\label{subsec:planar}

In this case, Eq. (\ref{eq:v}) becomes
\begin{equation}
  v = \frac{\sqrt{(1+ \rho_{r01}(z-z_0))^2-1}}{1+ \rho_{r01}(z-z_0)} c
\end{equation} 
where $\rho_{r01}$ is from Eq. (\ref{eq:density scale}).
The integral in Eq. (\ref{eq:t}) may be carried out to 
find, 
\begin{equation}
  t = \frac{\sqrt{(1+ \rho_{r01} (z-z_0))^2-1}}{\rho_{r01} c},
\end{equation}
which can be inverted to find $z(z_0,t)$ as
\begin{equation}
  z = z_0 + \frac{1}{\rho_{r01}}( f_1(z_0,t) - 1 ),\label{eq:z1D}
\end{equation}
where
\begin{equation}
  f_1(z_0,t) = \sqrt{1+\left(\rho_{r01} c t\right)^2},
\end{equation}
Taking the time derivative of Eq. (\ref{eq:z1D}), the velocity as a function of time becomes, 
\begin{equation}
  v = \frac{\rho_{r01}ct}{f(z_0,t)}c
\end{equation}
From Eq. (\ref{eq:den gen}), we find the density dynamics, 
\begin{equation}
  \rho_1(z;t) = \frac{\rho_{01}}{1 + \frac{d\rho_{r01}}{dz_0}\left[ {(ct)^2\over f_1(z_0,t)}  - \frac{(f_1(z_0,t)-1)}{\rho_{r01}^2} \right] }\label{eq:1D rho}
\end{equation}
where
\begin{equation}
  \frac{d\rho_{r01}}{dz_0} = {q\Sigma_{tot}\rho_{01} \over \epsilon_0 mc^2} = \frac{\rho_{01}}{l_{r01}} = \frac{\rho_{01}}{{\bar{\rho}}_{01}}\frac{\omega_{01}^2}{c^2}
\end{equation}

The non-relativistic limit occurs when $\rho_{r01} ct <<1$ or equivalently when $t<<t_x$ where
\begin{equation}
  t_x = \frac{1}{{\rho_{r01} c}}\label{eq:1D timescale},
\end{equation}  
and in this limit the expressions above reduce to the 
known results, i.e. $z = z_0 + qE_{01} t^2/2m$ and 
\begin{equation}
  \rho_{1NR}(z;t) = {\rho_{01}\over 1+{q \Sigma_{tot}\rho_{01}t^2\over 2\epsilon_0m} } = {\rho_0 \over 1 + \frac{1}{2} \frac{\rho_0}{{\bar{\rho}}_{01}}(\omega_{01}t)^2}\label{eq:1D ev non}
\end{equation}
where $\rho_{1NR}(z;t)$ is the density in the non-relativistic limit and 
$\omega_{01}$ is the plasma frequency defined in Eq. (\ref{eq:omega01}). 
\cite{Reed:2006_short_pulse_theory,Zerbe:2018_coulomb_dynamics}.    

\begin{figure*}
  \begin{center}
  \subfloat[]{\includegraphics[width=0.45\textwidth]{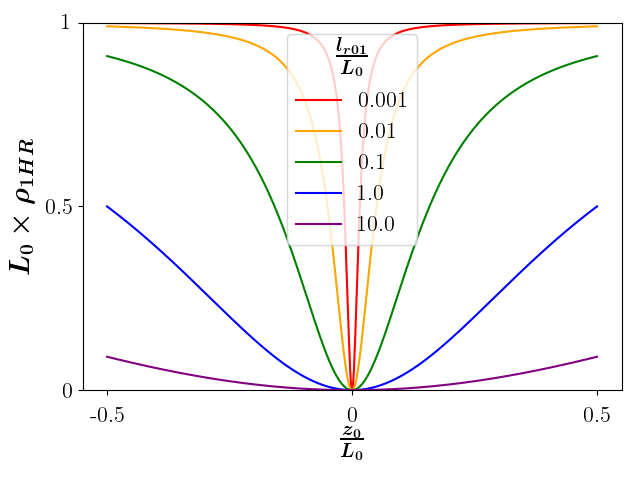}\label{fig:1d asym: unif}}
  \subfloat[]{\includegraphics[width=0.45\textwidth]{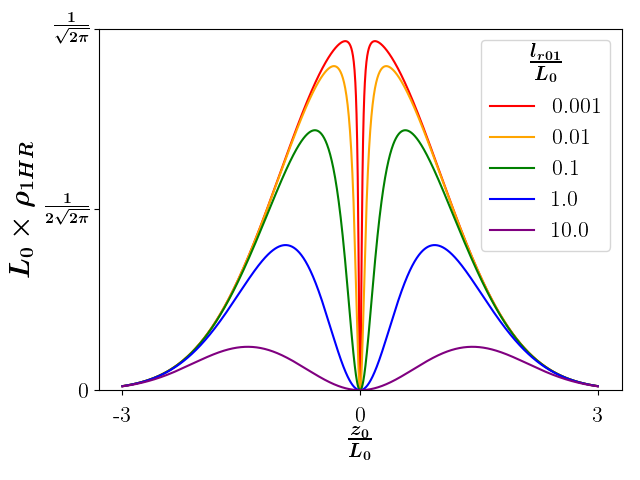}\label{fig:1d asym:Gaussian} }
  \end{center}
  
\caption{\label{fig:1d asym} Shape of the planar symmetric
asymptotic density for (a.) uniform and (b.) Gaussian initial
distributions.  $L_0$ represents the initial width of the distribution,
$l_{r01} =  \frac{\epsilon_0 mc^2}{q\Sigma_{tot}}$ is the length scale
associated with the density of the particles, and $\Sigma_{tot}$ is the charge per unit
area of the distribution as described in the text.  Notice that these graphs are independent
of the exact choices of $L_0$.  Further 
notice the quadratic like behavior in the middle as well as at large values
of $\frac{l_{r01}}{L_0}$ for the uniform distribution.  Finally note the fact that the distribution approaches
the original distribution at its maximum value when $\frac{l_{r01}}{L_0}$ is small.  What is not
displayed is that the maximum peak is proportional to $\frac{1}{l_{r01}}$ when 
$\frac{l_{r01}}{L_0}$ is large.}
\end{figure*}

The highly relativistic limit is when $\rho_{r01} ct >>1$ 
or equivalently $t>> t_x$.  Note that this second inequality implies that any 
point in the distribution except the center point at $z_0 = 0$ becomes highly relativistic
for sufficient time; this is part of the nature of the planar symmetries, and we find similar
nature for the cylindrical symmetries below.
In this limit, we find,
\begin{equation}
  z \rightarrow z_0 \pm ct
\end{equation}
where the sign of the luminal velocity is determined by on which side
of $z_0 = 0$ the particle originated
and 
\begin{align}
  \rho_{1HR}(z;t) \rightarrow \frac{\rho_{01}\rho_{r01}}{\rho_{r01}  + \frac{ \rho_{01}}{P_{01}}},
\label{eq:rel asym}
\end{align}
or equivalently
\begin{align}
  \frac{1}{\rho_{1HR}(z;t)} \rightarrow \frac{1}{\rho_{01}} + \frac{1}{P_{01} \rho_{r01}}
\end{align}
where $\rho_{1HR}(z;t)$ is the density distribution in the highly relativistic limit.  The interpretation 
of this result is interesting.   First, the majority of the distribution essentially becomes 
two pulses traveling at near luminal speeds away from one another. 
Second, as the particles within the distribution
reach luminal speeds, the density no longer significantly changes as 
the particles propagate to the left or right; that is, the density evolves toward an 
``asymptotic density" determined by Eq. (\ref{eq:rel asym}).  If $\rho_{r01} << \rho_{01}$, then
$\rho_{HR} \to \frac{(P_{01})^2}{l_{r01}}$; however, if $\rho_{01} >> \rho_{r01}$, then on the
edges $\rho_{HR} \to \rho_{01}$ whereas as you go further in the distribution transitions to
$\frac{(P_{01})^2}{l_{r01}}$.  This behavior for the uniform and Gaussian distributions for various
ratios of $\frac{l_{r01}}{L_0}$, where $L_0$ indicates the original width, 
may be seen in Fig. \ref{fig:1d asym}.

Analytically, for the case of 
an initial uniform distribution, $\rho_{01} = \frac{1}{L_0}$ and 
$P_{0} = \frac{2 z_0}{L_0} = 2 z_0 \rho_{01}$ where $L_0$ is the initial
width of the distribution.  In this case, 
\begin{equation}
  \rho_{1HR}(z) \to \frac{(2 z_0)^2}{(2 z_0)^2 L_0 + l_{r01} L_0^2}\label{eq:rel asym unif}.   
\end{equation}
Thus, the shape of this asymptotic distribution is
determined entirely by the length scale, $l_{r01}$, and the initial width, $L_0$.
For any point $z_0 << l_{r01}$, including the entire distribution
if $L_0 <<  l_{r01}$, this asymptotic density is essentially 
parabolic with zero density at the center and $\frac{1}{l_{r01}}$ at the edge.  
This case can be seen in Fig. \ref{fig:uniform 1d ev: Ea=0}.
For extremely dense distributions where 
$L_0 >> l_{r01}$, the asymptotic density at the edges approaches
the original density, $\rho_{01}$.  There is also a period of transition between the parabolic and
original density when the length scale is much smaller than the original width.  Both 
asymptotic behaviors can be
seen in Fig. \ref{fig:1d asym} for both the uniform and Gaussian cases. 
 
\begin{figure*}
  \begin{center}
  \subfloat[uniform]{\includegraphics[width=0.4\textwidth]{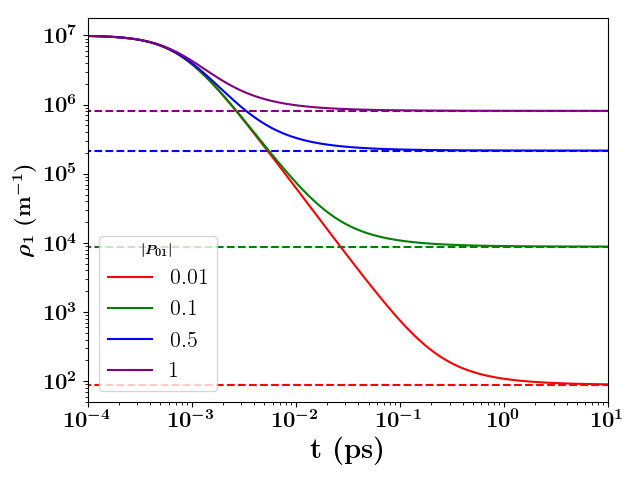}}
  \subfloat[Gaussian]{\includegraphics[width=0.4\textwidth]{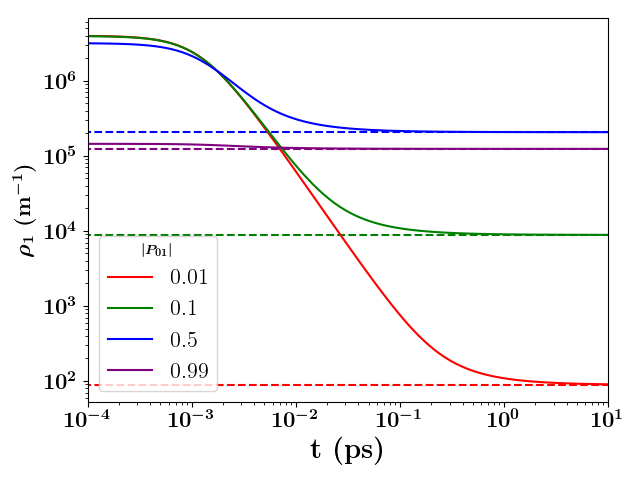}}
\end{center}
\caption{\label{fig:density vs t} Theoretical density evolution (solid lines)
of specific distribution points in the free expansion
of the bunch demonstrating the origin of the 
relativistic shock.  The dotted lines indicate the asymptotic value determined by Eq. (\ref{eq:rel asym}).
The points correspond to locations that symmetrically contain 
approximately 1\%, 10\%, 50\%, and either 100\% (uniform) or 99\% (Gaussian) 
of the distribution as indicated by their $P_{01}$ value.
The time scales of all points inversely correlate with their location in the distribution (see 
Eq. (\ref{eq:1D timescale})).  
For the uniform distribution (a),
all points start at the same density but converge to different asymptotes according to the inverse relationship
between position and time scale.  This leads to a parabolic distribution as seen in
Eq. (\ref{eq:rel asym unif}) as $L_0 <<  l_{r01}$ here.  
For the Gaussian distribution (b), 
the fact that the outer points have lower initial densities leads to density trajectories crossing
indicating the formation of the density peak that goes both up and down in contrast to the sharp
peak seen in the uniform case.}
\end{figure*}

\begin{figure*}
  \begin{center}
  \subfloat[]{\includegraphics[width=0.45\textwidth]{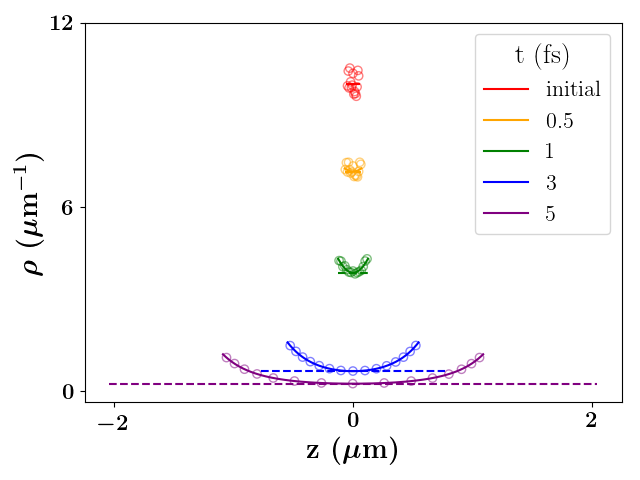}\label{fig:uniform 1d ev: Ea=0}}
  \subfloat[]{\includegraphics[width=0.45\textwidth]{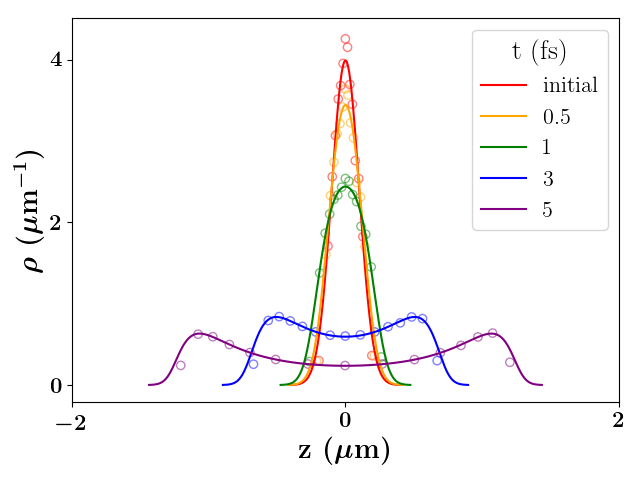}\label{fig:Gaussian 1d ev:Ea=0} }
  \end{center}
  
\caption{\label{fig:1d ev} Theoretical predictions (solid line) and $M$-shell simulations (hollow dots)
of the density at 4 different times with different initial conditions.  Notice the offset of the $0$ value on the
y-axis.  
The simulation had either (a) $L = 0.1 ~ \mu$m or (b) $\sigma_L = 0.1  ~ \mu$m, 
$\Sigma_{tot} = 10^8$ electrons per $\mu$m${}^2$, and
$M = 10,000$.  The extremely high density was chosen as this density does not require significant
expansion of the bunch before the onset of relativistic effects.
The dotted
lines in (a.) represents the non-relativistic prediction of the uniform distribution, and the
purple dotted line extends beyond the ends of the $x$-axis but are not shown to keep
the scale of expansion closer to what is seen relativistically; notice that
the dotted line is tangent to the relativistic distribution at the same time at $r = 0$. 
The deviation of the theory above the dotted line and toward the center of the bunch
indicates density freezing due to electrons obtaining luminal speeds whereas the non-relativistic
particles become super-luminal.  Simulations were
accomplished by randomly sampling $M$ macro-particles and advancing them according to 
the relativistic position equations derived in the text.}
\end{figure*}  

The mechanism for the relativistic peak emergence may be further seen in 
Fig. \ref{fig:density vs t} where the density associated with different Lagrangian particles within
the distribution are tracked and shown to asymptote to various density values predicted by
Eq. (\ref{eq:rel asym}).  One way to 
describe this mechanism is to notice that all particles, excepting the center particle, in a planar model 
will asymptote to the speed of light.  As the density is physically smooth, the 
particles' velocities in the 
neighborhood of the Lagrangian particle asymptote similarly to the speed of light.  In other words, the relative
velocity of the particles in the Lagrangian particle's neighborhood asymptotes to zero, and
the particles cease to spread in the $z$ dimension.  
As the $z$ dimension is the only dimension in which the density is spreading
in the planar model, this is the same as freezing the density to a constant value -- an asymptote.  
Moreover, as particles toward the edge of the distribution have larger accelerations,
these particles asymptote earlier than particles farther in.  These
differences in ``freezing'' time result in the middle of the distribution expanding, and becoming
less dense, before the onset of the relativistic regime.  Coupled with the initial distribution,
this results in the formation of density peaks toward the edge of the distribution,
as is seen in both the uniform and Gaussian distributions in Fig. \ref{fig:1d ev}.
In the non-relativistic limit, 
there is no Coulomb shock in planar bunches with cold initial conditions; while in the relativistic limit a 
strong shock emerges and an initial bunch described by either uniform or a Gaussian density distribution 
evolves to a two peak structure with one bunch moving to the right and the other to the left 
(see Fig. \ref{fig:1d ev}).

Also apparent in Fig. \ref{fig:1d ev} is the fact that stochastic effects are initially strong in 
simulated density profiles.  However, at long times the theoretical density
and simulated density agree well.  This is a real effect.  
Specifically, consider the inter-particle distance between
the $i^{th}$ and $(i+1)^{th}$ shells denoted as $d_i$.  For a uniform distribution, order statistics tells us that 
$d_i(0) = \frac{L}{M+1} + \epsilon$ where $L$ is the total width of the distribution, $M$ is the total number
of shells, and $\epsilon$ is a stochastic factor roughly of the size $ \frac{L}{M+1}$.  Thus, due to stochastic
fluctuations, we'd expect some sheets to be bunched together giving a higher local density than 
the average and likewise other sheets to be further apart giving a lower local density than the average.
This is precisely what is seen with the initial distribution in Fig. \ref{fig:1d ev}.  However, as these sheets
evolve, the relative non-relativistic acceleration is $\frac{2 q E_{T1}}{m M}$, so $d_i(t) = d_i(0) + \frac{q E_{T1}}{m M} t^2$.  Given sufficient time, 
$t  >> \sqrt{\frac{m L}{q E_{T1}}}$, $d_i(t) \approx \frac{q E_{T1}}{m M} t^2$.  That is, the inter-particle
distance (and hence the distribution) is dominated by the space-charge effect and converges to the
space-charge predicted distribution everywhere.  Of course, if the bunch enters the relativistic 
regime prior to this smoothing, the stochastic effects will be preserved.  We will see such behavior once
we add an extraction field, but such behavior requires extremely dense bunches that may not be
physically possible in 
free expansion experiments.

\subsection{Cylindrical symmetry}\label{subsec:cyl}
Now we consider the expansion of an initially cold charged particle cloud with cylindrical symmetry.    
In this case, Eq. (\ref{eq:v}) becomes
\begin{align}
  v &= \frac{\sqrt{2}\zeta y  \sqrt{1 + \zeta^2 y^2 }}{1 + 2 \zeta^2 y^2}c\label{eq:2D rel v}
\end{align}
where $y^2 = \ln\left(\frac{r}{r_0}\right)$, 
 $\zeta^2 = \frac{r_0\rho_{r02}}{2} = \frac{r_0 P_{02}}{2 l_{r02}} = \frac{r_0^2 \omega_{02}^2}{4 c^2}$, with
$\rho_{r02}$
coming from Eq. (\ref{eq:density scale}).  As $l_{r02} \propto r_0$, it should be apparent that
$\zeta$'s dependence on $r_0$ is completed determined by $P_{02}(r_0)$.

From Eq. (\ref{eq:2D rel v}) and (\ref{eq:t}), we find the implicit relation between time and radial position through the 
integral,
\begin{align}
  t &= \frac{2}{\omega_{02}}\int_{0}^{\sqrt{\ln\left(\frac{\tilde{r}}{r_0}\right)}} \frac{1 + 2\zeta^2 y^2}{\sqrt{1 +  \zeta^2 y^2}} e^{y^2} dy.
\end{align}
To make the connection with previous work, we introduce a generalized Dawson function $\mathcal{F}$ 
through the definition,
\begin{align}
  \mathcal{F}(g,x) = e^{-x^2}  \int_{0}^{x} g(\zeta,z)~ e^{z^2}dz\label{eq:mod Dawson def}
\end{align}
where $\zeta$ can be written as a function of $x$.  Thus the time-spatial relation may be
expressed as
\begin{align}
  t &= \frac{2}{\omega_{02}} \frac{r}{r_0} \mathcal{F}\left(g(\zeta,y), y\right) \label{eq:2D time int}
\end{align}  
where
\begin{align}
  g(\zeta,y) = \frac{1 + 2 \zeta^2y^2}{\sqrt{1 + \zeta^2 y^2}}
\end{align}
When  $g(\zeta,y) = 1$,  we reproduce the Dawson function, $F(x) = \mathcal{F}(1,x)$.
Specifically when we are in the non-relativistic regime, we have $2 \zeta y << 1$ and 
$g(\zeta,y) \approx 1$, so 
Eq. (\ref{eq:2D time int}) reduces to
\begin{align}
  t &\approx \frac{2}{\omega_{02}} \frac{r}{r_0} F(y) \label{eq:2D time}
\end{align}
which is the result we derived previously in the non-relativistic case.  We can write down the derivative 
of the generalized Dawson function by applying the Leibniz rule
\begin{align}
  \frac{d \mathcal{F}}{dx} &= - 2 x \mathcal{F}(g,x) + g(\zeta,x) + \mathcal{F}\left(\frac{\partial g}{\partial \zeta},x\right) \frac{d\zeta}{dx} \label{eq:mod Dawson deriv}
\end{align}
Note that in the non-relativistic limit,  
$g(\zeta,y) = 1$, and Eq. (\ref{eq:mod Dawson deriv}) reduces to the 
normal Dawson function derivative $ \frac{dF}{dx} = - 2 x F(x) +1$.

Following the same reasoning as our previous work\cite{Zerbe:2018_coulomb_dynamics},
 we can obtain an analytic form for the
time dependent density, i.e. the density evolution expression (see Eq. (\ref{eq:den gen})).    Evaluating  
$r' = \frac{dr}{dr_0}$ by taking a derivative 
of Eq. (\ref{eq:2D time int}) with respect to $r_0$, we find,
\begin{align}
  r' &= \frac{r}{r_0}\left(1 + \frac{2 y}{g(\zeta,y)} \left(
D_{02}\ \mathcal{F} - \frac{\rho_{02}}{{\bar{\rho}}_{02}} \zeta \mathcal{F}_\partial \right)\right)\label{eq:cyl drdr0}
\end{align}
where $\mathcal{F}$ is shorthand for $\mathcal{F}(g(\zeta,y),y)$
$\mathcal{F}_\partial$ is shorthand for
$\mathcal{F}\left(\frac{\partial g}{\partial \zeta},y\right)$,
and $D_{02}$ is from Eq. (\ref{eq:D_{0d}}).
Note $D_{02}$  measures the deviation from a uniform cylindrically-symmetric distribution, and for the 
uniform cylindrically-symmetric distribution case it is zero for all values of $r_0$ where $\rho_0$ is not $0$. 

From the above analysis, the density evolution is found to be, 
\begin{align}
  \rho_2(r;t) &= 
                     \frac{r_0^2}{r^2} 
                     \frac{\rho_{02}}{  1 + \frac{2 y}{g(\zeta,y)} 
                     \left( D_{02}\ \mathcal{F} - \frac{\rho_{02}}{{\bar{\rho}}_{02}} \zeta \mathcal{F}_\partial\right)  }\label{eq:cyl rho}
\end{align}

\begin{figure*}
  \begin{center}
  \subfloat[]{\includegraphics[width=0.45\textwidth]{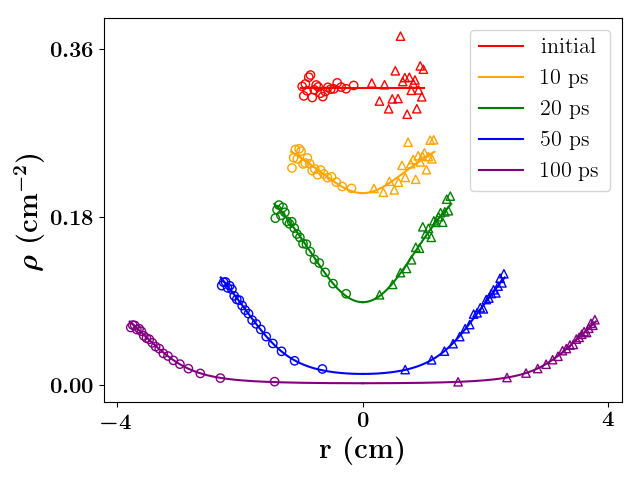}}
  \subfloat[]{\includegraphics[width=0.45\textwidth]{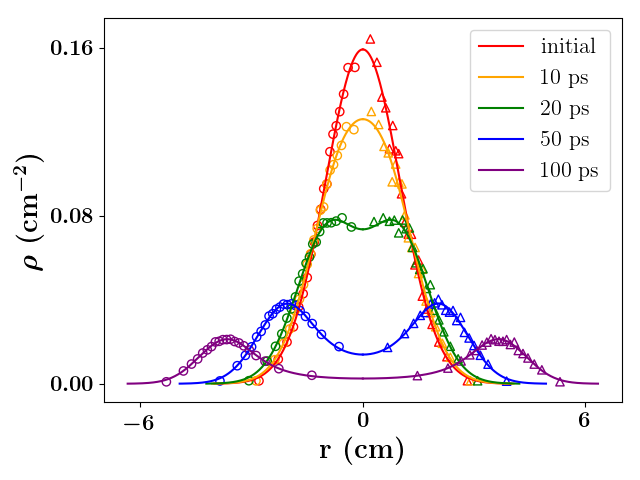}}
  \end{center}
\caption{\label{fig:2d ev} Theoretical predictions (solid line), cylindrical
$M$-shell simulations (hollow triangles--right), and
PIC simulations using an EM solver (hollow circles--left)
of the density at 5 different times for the cylindrically symmetric case.   
PIC simulations were analyzed using the middle 2m from a simulation of a 20m-long, 
3D distribution of electrons with periodic boundary
conditions on the z-axis.
The theory prediction is reflected about the origin as $r$ is 
strictly greater than 0
and the simulations are separated into the first and second quadrant for purposes of visualization.
Parameters were $N = 1 \times 10^{13}$  electrons/cm, $R_0=\sigma_{r,0} = 1~$cm, and $M = 50,000$.
Like the planar symmetric case, the density at the center continues to
decrease non-relativistically indicating that the density above the density at the center is due to 
relativistic effects; however, notice that
this peak continues to decrease instead of the evolution freezing at luminal speeds as seen in 
the planar symmetric case.}
\end{figure*}

In Fig. \ref{fig:2d ev}, 
we compare the predictions of Eq. (\ref{eq:cyl rho}) to simulations for both uniform and Gaussian
initial distributions.  We choose the initial radius and radial standard deviation, respectively, to be 1 cm
for $N = 1 \times 10^{13}$ electrons/cm.  We again simulate with Warp using the EM solver as well as
the 2D version of 
$M$-shell simulations.  For the $M$-shell simulations, the initial radius of the $M$ cylindrical shells 
are sampled and then evolved according to 
 Eq. (\ref{eq:2D rel v}) and Eq. (\ref{eq:2D time}) but with 
 $\omega_{02}$ replaced by $\sqrt{\frac{3 q \Lambda_{s} }{\pi r_{s,0}^2 m \epsilon_0}}$
 where $\Lambda_s$ is the charge 
 per unit length contained in the cylindrical shell and $r_{s,0}$ is the initial radius of the shell.
 As can be seen in Fig. \ref{fig:2d ev}, the theory and both simulations agree on the 
 evolution of both the uniform and non-uniform
 initial distributions.  Similar to the planar case, the initial variance about the predicted value can be seen
 to decrease as the simulations evolve.  Again, this indicates that the
 inter-shell distances are dominated by the space-charge effects resulting in the later simulations 
 having less statistical variation from the expected distribution.

In the non-relativistic regime, $2 \zeta y  << 1$;
$\mathcal{F} \to F$ and $\zeta \mathcal{F}_\partial \to 0$.
Thus Eq. (\ref{eq:cyl rho}) reduces to 
\begin{align}
  \rho_2(r;t) &= \frac{r_0^2}{r^2} \frac{\rho_{02}}{1 + 2 D_{02}\ yF(y)}\label{eq:cyl rho non}
\end{align}
which is the expression we found in our earlier, non-relativistic 
work\cite{Zerbe:2018_coulomb_dynamics}.  

Similar to the planar symmetric case, we are interested in the density the distribution 
should evolve under specific limits.  We were unable to analytically obtain a limit analogous
to the limit we found under planar symmetry
in Eq. (\ref{eq:rel asym}) as doing so requires evaluating the value of the modified
Dawson function as $\ln\left(\frac{r}{r_0}\right)$ goes to infinity.  
We were able to see the freezing of the dimension in 
the extremely dense limit
where $\zeta >> 1$ as we have shown in Appendix \ref{ap:high density} .  Specifically, the
evolution at the edge of the distribution can be approximated by 
$\rho_2(r;t) = \frac{r_0}{r} \rho_0$, which is the evolution of the uniform distribution
under non-relativistic conditions in one-dimension lower, i.e. 1D.  This situation is
analogous to the high density 1D case that causes the edges to essentially immediately
become relativistic likewise resulting in evolution of the uniform distribution
under non-relativistic conditions in one-dimension lower, i.e. 0D or constant.
However, this condition, $\zeta >> 1$, is analogous to the 1D case when the
entire distribution is essentially in the highly relativistic limit.
We will shortly show that even in this case, the spherically symmetric evolution can be
shown to freeze out a dimension; however, we believe that this freezing
happens for cylindrically symmetric distributions regardless of the size of $\zeta$. 

\subsection{Spherical symmetry}\label{subsec:sph}
Now we consider the expansion of an initially cold charged particle cloud with spherical symmetry.    
In this case, Eq. (\ref{eq:v}) becomes
\begin{align}
  v &=   2 \zeta x  \frac{\sqrt{g_1(x)}}{g_2(x)}c\label{eq:3D rel v}
\end{align}
where  $x^2 = 1 - \frac{r_0}{r}$,  $g_1(x) = 1 + \zeta^2x^2$,
$g_2(x) = 1 + 2\zeta^2x^2$,
$\zeta^2 =\frac{r_0 \rho_{r03}}{2} = \frac{r_0P_{03}}{2 l_{r03}} = \frac{r_0^2\omega_{03}^2}{6c^2}$, and
$\rho_{r03} $
is from Eq. (\ref{eq:density scale}).
As $l_{r03} \propto r_0^2$, it should be apparent that
$\zeta \propto r_0 P_{03}(r_0)$.

From Eq. (\ref{eq:2D rel v}) and (\ref{eq:t}), we find the implicit relation between time and radial position through the 
integral,
\begin{align}
  t &= \frac{ \sqrt{3/2} }{ g_1(1) \omega_{03} } \left( g_2(1)\frac{r}{r_0} x \sqrt{g_1(x)} + T(x)\right)\label{eq:3D time}
\end{align}
where $T(x) = \tanh^{-1}\left(\sqrt{\frac{g_1(1)}{g_1(x)}}x\right)$.  Note that
the $1$ inside the $g$ functions corresponds to $x$ at infinitely long times, i.e. 
$\lim_{\frac{r}{r_0} \to \infty} x = 1$, so $g_1(1) = 1 + \zeta^2$ and 
$g_2(1) = 1 + 2 \zeta^2$.
This expression is essentially the same expression as derived by Bychenkov and Kovalev, 
who first derived it
for the case of uniform initial density distributions \cite{Bychenkov:2011_relativistic}.  
Our expression differs only in the 
interpretation of $\omega_{03}$ as ours can be
dependent on $r_0$ whereas their $\omega_{03}$ is a constant, which is 
the correct interpretation for the uniform distribution.  This difference 
in interpretation allows us to treat
general initial distributions but requires additional consideration when determining
the derivative of Eq. (\ref{eq:3D time}) with respect to $r_0$ as 
$\omega_{03}' = \frac{3 \omega_{03}}{2 r_0}D_{03}$,
where $'\equiv\frac{d}{dr_0}$, with
the $d$'s in this last expression representing differentiation -- not dimensionality of the problem,
and $D_{03}$ is from Eq. (\ref{eq:D_{0d}}).

We follow the same reasoning as our previous work \cite{Zerbe:2018_coulomb_dynamics} 
in order to obtain
the density evolution expression.  After taking the derivative
of Eq. (\ref{eq:3D time}) with respect to $r_0$, we can solve for $r'$ giving
\begin{align}
  r' &= \frac{r}{r_0}\frac{1}{( g_1(1) )^2 g_2(x)}\left(p_1(x)
+ \frac{r_0}{r} p_2(x) T(x)\right)
\label{eq:3D drdr0}
\end{align}
where
\begin{align}
  p_1(x) &= g_1(1) + \frac{3}{2} D_{03} x^2 g_1(x) + 3 \zeta^4 D_{03}x^2\frac{r_0}{r}
\end{align}
and
\begin{align}
  p_2(x) &= \left(3\zeta^2 + 6 \zeta^2 D_{03} + \frac{3}{2} D_{03}\right) x \sqrt{\frac{g_1(x)}{g_1(1)}}
\end{align}
Plugging Eq. (\ref{eq:3D drdr0}) into Eq. (\ref{eq:den gen}) we obtain
the evolution of the density distribution 
\begin{align}
  \rho_3(r;t) = \frac{r_0^3}{r^3} \frac{( g_1(1) )^2 g_2(x) \rho_0}{p_1(x)
+ \frac{r_0}{r} p_2(x) T(x)}\label{eq:3D rho}
\end{align}

\begin{figure*}
  \begin{center}
  \subfloat[]{\includegraphics[width=0.45\textwidth]{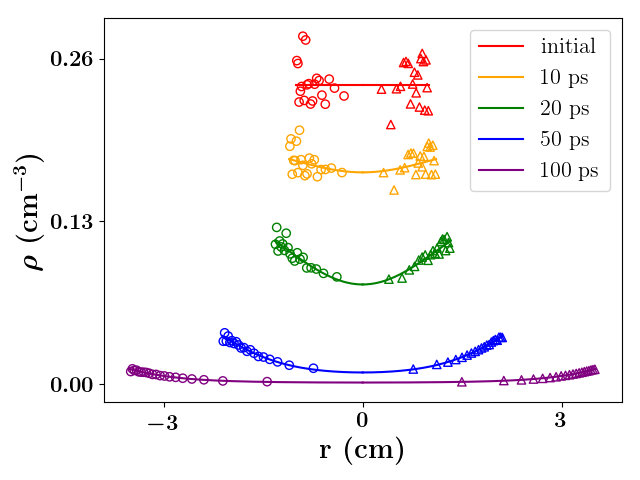}}
  \subfloat[]{\includegraphics[width=0.45\textwidth]{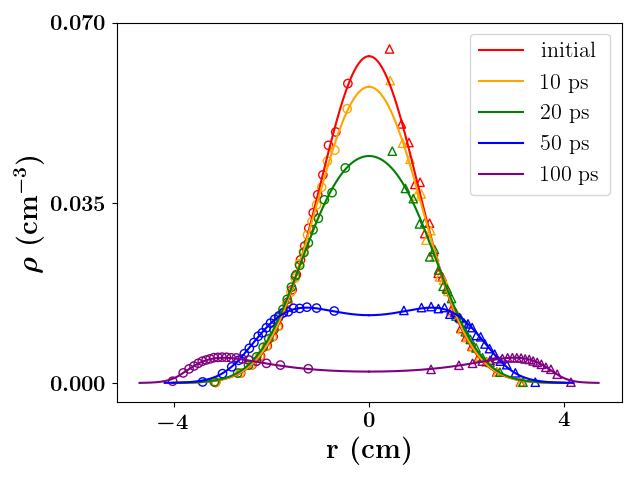}}
  \end{center}
  
\caption{\label{fig:3d ev} Theoretical predictions (solid line), spherical 
$M$-shell simulations (hollow triangles--right), and
PIC simulations using an EM solver (hollow circles--left)
of the density at 5 different times for the spherically symmetric case.   
The theory is reflected about the origin as $r$ is strictly greater than 0
and the simulations are separated into the first and second quadrant for purposes of visualization.
Parameters were $N = 1 \times 10^{13}$  electrons, $R_0=\sigma_{r,0} = 1~$cm, and $M = 5000$.
Like the planar and cylindrical symmetric cases, the density at the center continues to
decrease non-relativistically indicating that the density above the density at the center is due to 
relativistic effects; similar to the cylindrical symmetric case,
the relativistic density continues to evolve.  However, the evolution of the peak
decreases much faster than the cylindrical case --- which is discussed in detail in the text.}
\end{figure*}

In Fig. 5, we compare the prediction of Eq. (\ref{eq:3D rho}) to simulations for both a uniform and Gaussian
initial distributions of $N = 1 \times 10^{13}$ electrons.  
The simulation have $R = 1$ cm (uniform) or $\sigma_r = 1$ cm (Gaussian).  
We again simulate with Warp using the EM solver and as well as
the 3D version of 
$M$-shell simulations.  For the $M$-shell simulations, the initial radius of the $M$ spherical shells 
are sampled and then evolved according to 
 Eq. (\ref{eq:3D rel v}) and Eq. (\ref{eq:3D time}) but with 
 $\omega_{03}$ replaced by $\sqrt{\frac{3 q Q_{s} }{ 4 \pi r_{s,0}^3 m \epsilon_0}}$
 where $Q_s$ is the charge contained in the shell and $r_{s,0}$ is the initial sampled radius of the shell.
 As can be seen in Fig. \ref{fig:3d ev}, the theory captures the evolution of both the uniform and non-uniform
 initial distributions.  Similar to both the planar and cylindrical cases, the initial variance around the theoretical 
 value primarily seen in the
 uniform distribution decreases as the distribution expands.

For further validation, we compare Eq.  (\ref{eq:3D rho}) to the expression derived by Bychenkov
and Kovalev.  Their expression detailed the relativistic density evolution 
for the uniform distribution\cite{Bychenkov:2011_relativistic}, $\rho_{unif}(r;t)$,
which should be equivalent to our expression when $D_{03} = 0$.  
In this case, Eq. (\ref{eq:3D rho}) reduces to
\begin{align}
  \rho_{3unif}(r;t) &= \frac{r_0^3}{r^3} \frac{(g_1(1))^2 g_2(x) \rho_0}{g_1(1) + \frac{r_0}{r} 3\zeta^2 x  \sqrt{\frac{g_1(x)}{g_1(1)}} T(x)} \label{eq:3D drdr0 unif}
\end{align}
This expression for the density evolution for uniform initial conditions
 is identical to the expression published in the English translation of Bychenkov and 
Kovalev except for an obvious typo in that work\cite{Bychenkov:2011_relativistic}.  

Next, we compare this expression to our previous, non-relativistic expression.
In the non-relativistic regime $2 \zeta^2 << 1$.  Unlike the planar and
cylindrical cases, the spherical model need never enter the relativistic regime 
and therefore this model may be relevant for all time.  In this
non-relativistic regime, Eq.  (\ref{eq:3D rho}) reduces to
\begin{align}
  \rho_{3NR}(r;t) &= \frac{r_0^3}{r^3} \frac{\rho_0}{1 + \frac{3}{2} D_{03} \left(x^2 + \frac{r_0}{r} x \tanh^{-1} x\right)}
\end{align}
which is identical to the non-relativistic expression we previously derived but with 
$D_{03} = D$ in our previous notation\cite{Zerbe:2018_coulomb_dynamics}.

\begin{figure*}
  \begin{center}
  \subfloat[]{\includegraphics[width=0.45\textwidth]{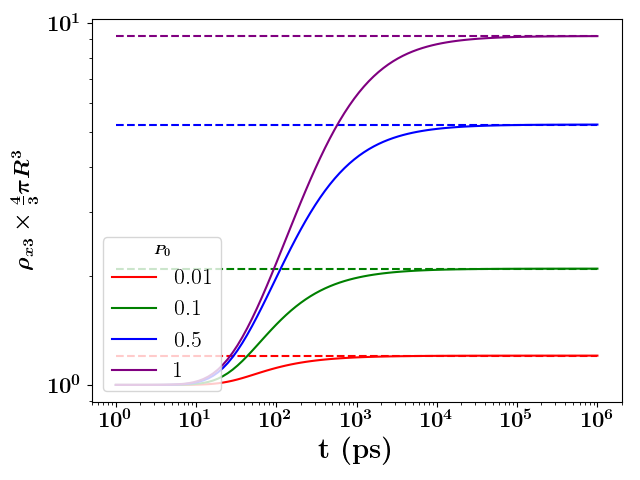}}
  \subfloat[]{\includegraphics[width=0.45\textwidth]{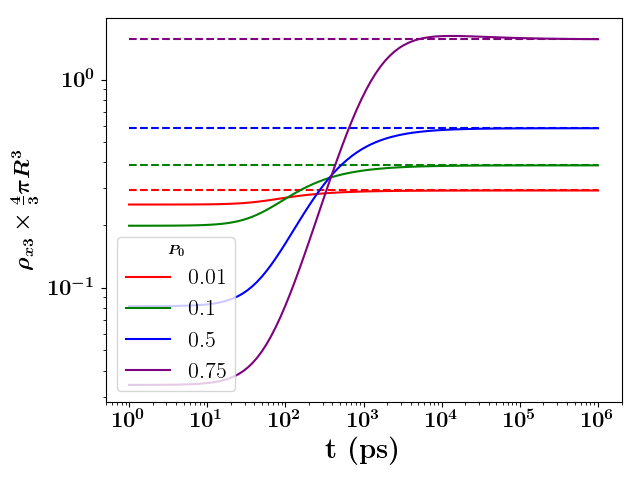}}
  \end{center}
  
\caption{\label{fig:3d time asym} The time evolution (solid lines) of $\rho_{x3}$, defined in the text, 
for a distribution with 
$1.602 mC$ in (a) an initially uniform sphere of width 1 mm and (b) an initially Gaussian 
sphere with standard deviation of 1mm.  The dashed lines indicate the corresponding 
highly-relativistic limit of $\rho_{x3}$ obtained analytically.  
Again, the formation of the density peak from these
relativistic considerations is apparent in the graphs.  Similar to how the planar symmetric density freezes,
$\rho_{x3}$ can be seen to asymptote; however, this is due to the Lagrangian particle reaching 
their terminal velocity, the difference of which is relativistically contracted, as described in the text.
}
\end{figure*}

\begin{figure*}
  \begin{center}
  \subfloat[]{\includegraphics[width=0.45\textwidth]{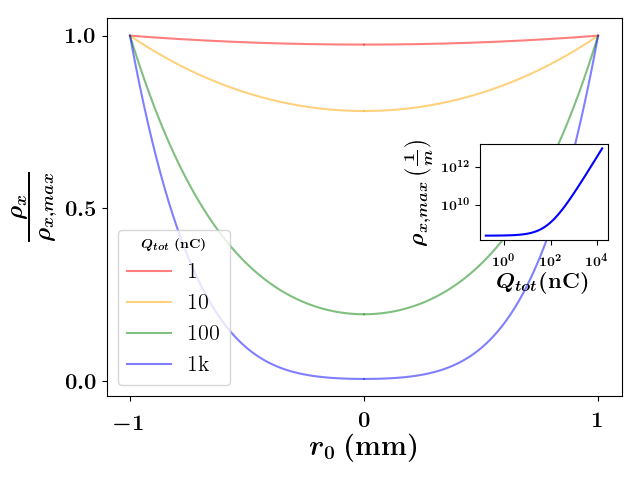}}
  \subfloat[]{\includegraphics[width=0.45\textwidth]{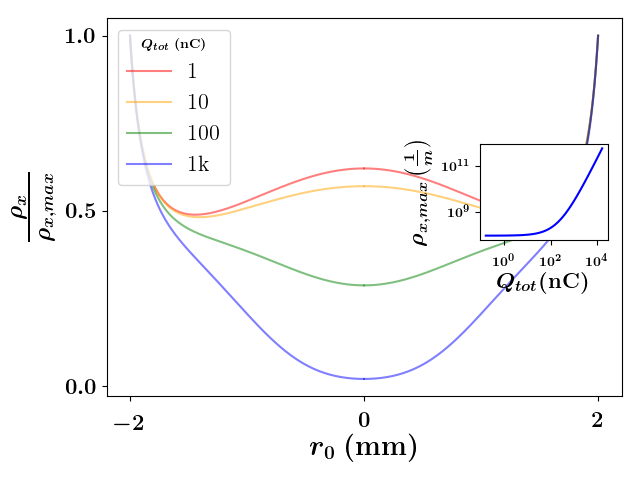}}
  \end{center}
  
\caption{\label{fig:3d asym} The shape of $\frac{\rho_{x3}}{\rho_{x3,max}}$ 
for various values of $Q_{tot}$ in (a) an 
initially uniform distribution with radius of 1 mm and (b) an initially Gaussian distribution with
standard deviation of 1 mm
as a function of the initial radial coordinate.  This, as well as the inset figure describing the 
growth of the maximum value of $\rho_{x3}$, show the onset of the relativistic regime corresponding
to roughly $24 \frac{C}{m}$.
The main graph also shows that the peaks sharpen as the density increases.
For the Gaussian case, this distribution diverges near $r_0 = 2$.  This divergence is an indication
that the laminar fluid assumption is being violated, and therefore the non-uniform 
asymptote for $\rho_{x3}$ should be taken as a gross approximation; nonetheless,
the expression derived in the text do capture the shape of freely-expanding relativistic Gaussian
distribution in the simulation presented in Fig. \ref{fig:3d ev}.  L}
\end{figure*}

Again we would like to analyze specific limits of the density evolution; fortunately,
under spherical symmetry we can analyze the long time limit.  In Appendix \ref{ap:long time} we
show that 
\begin{align}
  \lim_{\frac{r}{r_0} \to \infty} \rho_3 &= \frac{r_0^3}{r^3} \rho_{x3}(r_0)\label{eq:intro rhox}
\end{align}
where 
$\rho_{x3}(r_0) = \frac{1 + 3 \zeta^2 + 2 \zeta^4}{1 + \frac{3}{2} D_{03}} \rho_{03}$ is entirely determined by the initial conditions.
Notice, the pre-factor in  $  \frac{1 + 3 \zeta^2 + 2 \zeta^4}{1 + \frac{3}{2} D_{03}}$ 
is essentially $1$ in the center where 
$\zeta \approx 0$ and $D_{03} \approx 0$, but that this value increases as $r_0$ increases.
The time evolution of $\rho_{x3}$ and the predicted asymptote for this quantity can be seen in Fig.
\ref{fig:3d time asym}.
For the uniform distribution, the increase in $\rho_{x3}$ as a function of $r_0$ is quartic as $D_{03} = 0$ for all values of $r_0$.  
In real distributions, though, there should be a value for $r_0$ where $D_{03} = -\frac{2}{3}$, and
we see that $\rho_{x3}$ has a zero in the denominator.  This violates the assumptions made
in the derivation of $\rho_{x3}$, and inspection of Appendix \ref{ap:long time} shows that 
$r'$ becomes $0$ in the locality of $D_{03} = -\frac{2}{3}$ suggesting a violation of the 
laminar fluid assumption.  
For the Gaussian distribution, roughly 80\% of the distribution is contained within the
radius where $D_{03}(r_0) = -\frac{2}{3}$ suggesting that at least the majority of the distribution is
captured by this theory.  Furthermore, the
the precise shape for $\rho_{x3}(r_0)$ 
for a uniform and Gaussian distribution may be seen in Fig. \ref{fig:3d asym}; however,
in the 1D case, $\rho_{01}$ truly asymptotes whereas here $\rho_{03}$ continues to decrease eventually 
with the uniform-like behavior of $\frac{r_0^3}{r^3}$.  This difference is largely due to the fact that
all particles asymptote to the same velocity, $c$, in the planar case but different 
velocities in the spherical case.

This analysis leads to the second limit of $\lim_{\zeta \to \infty}$, an unphysical limit,
analogous to the 1D and cylindrically-symmetric cases where we saw the density at the 
edge lose dimensionality.
Likewise, in the spherical-symmetric case we show in Appendix \ref{ap:high density} that
\begin{align}
  \lim_{\zeta \to \infty}\rho_3(r;t)  = \frac{r_0^2}{r^2} \rho_{03}
\end{align}
which is again the uniform density evolution of a symmetric distribution in one dimension less than
the one being considered.  While appearing unphysical, this does have some 
physical significance, though.  This suggests that such distribution evolve toward
$\rho_x(r_0)$ in such a way that the factor in the denominator exactly cancels out the factor of 
$\frac{r_0}{r}$, i.e.
$r' \approx 1$ early on.  However, as time progresses, the evolution shifts toward
the decay of the uniform distribution in the appropriate dimension.

\section{Extraction field in the planar model}\label{sec:Ea}

\begin{figure*}[tbh]
  \begin{center}
  \subfloat[uniform]{\includegraphics[width=0.4\textwidth]{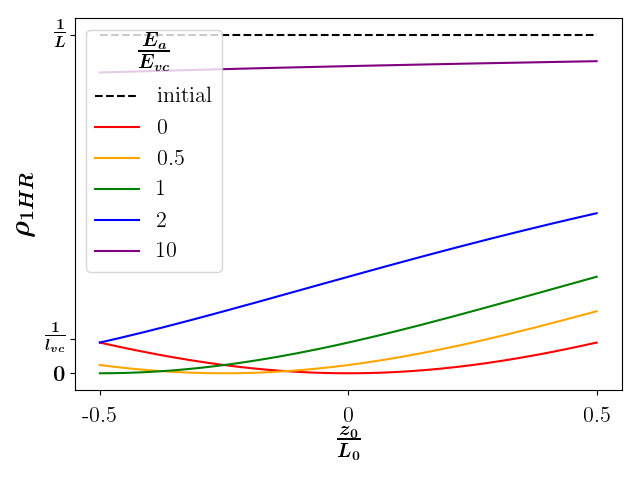}}
  \subfloat[Gaussian]{\includegraphics[width=0.4\textwidth]{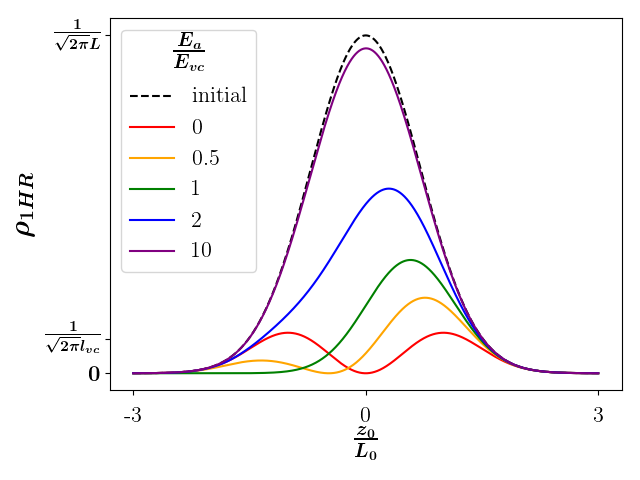}}
\end{center}
\caption{\label{fig:asymptotes} Theoretical planar symmetric asymptotic distributions (solid lines) 
plotted against the initial position
for various applied fields, measured in multiples of $E_{T1} = \frac{\Sigma_{tot}}{2 \epsilon}$.  The black
dotted line indicates the initial distribution.  Notice that in both cases, the actual distribution is located
at $z(t)$ at time $t$, but as this has a one-to-one relation with $z_0$, we use the initial location for the 
sake of comparison.}
\end{figure*}

It is straightforward to introduce an extraction field in the planar model, and this is relevant to 
dynamics of electron density distributions in the pancake bunches used in 
ultrafast electron microscopy.  
The equations for this case are identical to the equations derived for the planar model in the absence
of an extraction field with the single replacement
\begin{align} 
  P_{01}&\rightarrow P_{01} + \frac{E_a}{E_{T1}}\label{eq:Eeff}
\end{align}
or equivalently
\begin{align}
  \rho_{r01} &\rightarrow \rho_{r01} + \rho_a\label{eq:rel transform}
\end{align}
where $\rho_{r01}$ is from Eq. (\ref{eq:density scale}) and $\rho_a = \frac{eE_a}{mc^2}$, which
also can be interpreted as a new length scale, $l_a = \frac{mc^2}{eE_a}$, associated with
the extraction field.
We choose the applied field to be in the positive $z$ direction. 
For applied fields, $E_a$, with $E_a>E_{T1}$ ($l_a < l_{r01}$) the applied field 
is sufficiently strong to overcome the space charge field throughout the bunch, hence
accelerating all of the particles in the same direction.   
For smaller applied fields, 
$E_a < E_{T1}$ ($l_a > l_{r01}$),  
particles in the negative $z$ regions of the initial charge distribution may 
experience a stronger intrinsic space charge field than can be overcome by the applied field. 
In this case the initial distribution breaks up into two bunches moving in opposite directions.
This is the virtual cathode limit defined by Valfells et al.\cite{Valfells:2002_vc_limit}.  
We also point out that $l_{r01}$ corresponds to
the length scale of this limit. 
  The fraction of charge in the bunch that moves in the positive and negative $z$ direction is simply
$ \frac{1}{2} \left(1 \pm \frac{E_a}{E_{T1}}\right)$, respectively .   
The form of the two bunches is given by Eq.(\ref{eq:1D rho}), 
with the substitution given in Eq. (\ref{eq:Eeff}).   Taking the relativistic limit of this expression, we 
obtain the asymptotic form of the two bunches
\begin{align}
  \rho_{1HR}(z;t) &\rightarrow \frac{\rho_{01}}{\frac{1}{l_{r01}}\left(P_{01} + \frac{E_a}{E_{T1}}\right)^2  + \rho_{01}} \left(P_{01} + \frac{E_a}{E_{T1}}\right)^2 \label{eq:Ea rel asym}
\end{align} 
and the asymptotic form for the uniform and Gaussian distributions for various applied fields 
are demonstrated in Fig. \ref{fig:asymptotes}.  Eq. (\ref{eq:Ea rel asym}) can be written in terms of
the plasma period, $\omega_{01}$, and other terms, but we find that both the applied field scale, 
$E_{T1}$, and the associated length scale, $l_{r01}$, are more apparent in this formulation.

For the case of a uniform initial distribution $\rho_{01} = \frac{1}{2L}$ on the domain $[-L,L]$,
the field of a particle at position $z_0$ is given by
\begin{equation}
  E_{01}(z_0) + E_a = E_{T1}  \frac{z_0}{L} + E_a
\end{equation}
Setting the total field to zero gives the point at which the 
pulse breaks up into two pulses,
\begin{equation}
z_x = -\frac{E_a}{E_{T1}} L
\end{equation}
Notice that as $E_a$ goes above $E_{T1}$, $z_x$ goes below $-L$
indicating that there is no split in the pulse consistent with prior analysis.  As long as  $E_a < E_{T1}$,
the peak density of each pulse after it has gone relativistic can be calculated from 
Eq. (\ref{eq:Ea rel asym}) and is 
\begin{align*}
  \rho_{1right/left} &= \frac{1}{2L  + l_{r01} \left(\frac{E_a}{E_{T1}} \pm 1\right)^{-2}}
\end{align*}
again with the positive, rightward pulse corresponding to the $+$ and the negative, leftward pulse
corresponding to the $-$.

\begin{figure*}[tbh]
  \begin{center}
  \subfloat[$E_a = 0.5 ~ E_{T1}$]{\includegraphics[width=0.45\textwidth]{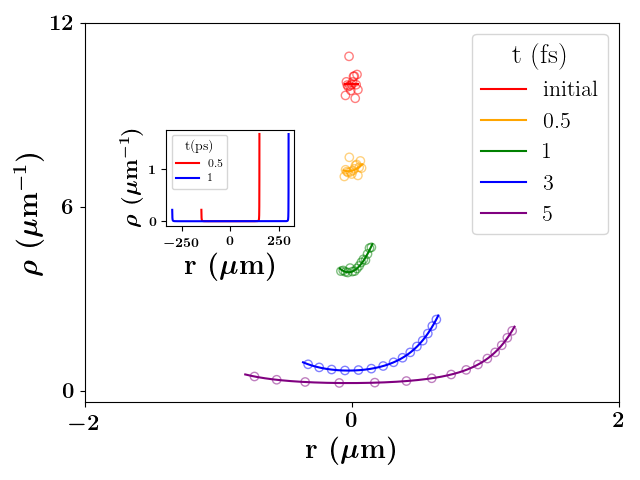}}
  \subfloat[$E_a = E_{T1}$]{\includegraphics[width=0.45\textwidth]{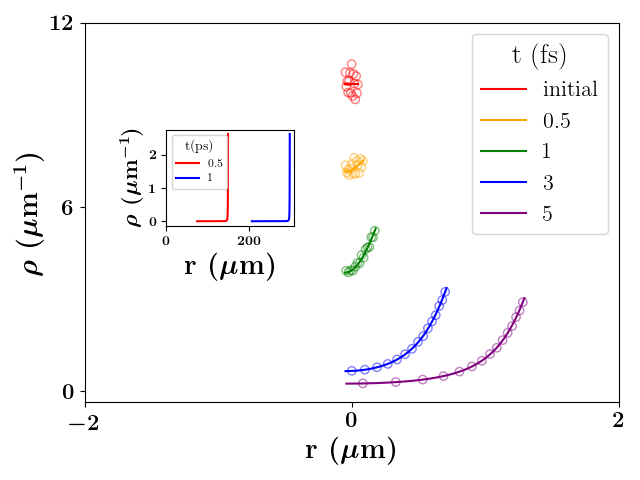}}\\
  \subfloat[$E_a = 2 ~ E_{T1}$]{\includegraphics[width=0.45\textwidth]{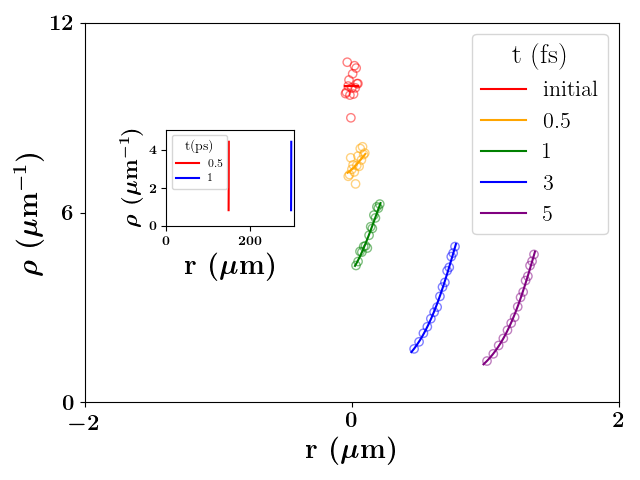}}
  \subfloat[$E_a = 10 ~ E_{T1}$]{\includegraphics[width=0.45\textwidth]{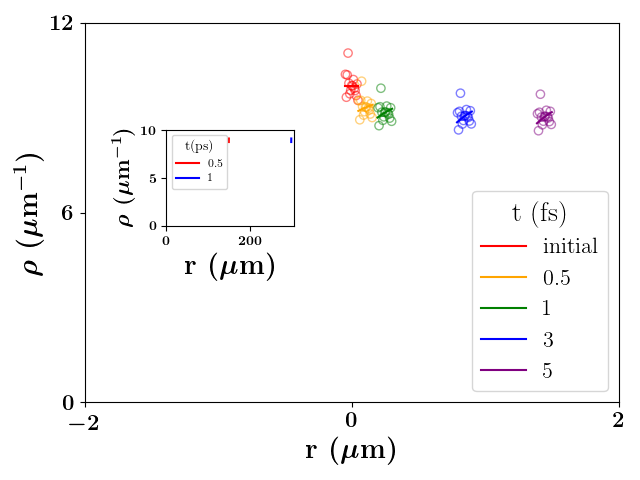}\label{fig:uniform 1d ev: Ea=10}}
  \end{center}
    
\caption{\label{fig:uniform 1d ev} Theoretical predictions (solid line) and $M$-shell 
simulations (hollow dots)
of the planar symmetric density at 4 different times with different applied extraction fields.  
The simulation had $L = 0.1 ~ \mu$m and $\Sigma_{tot} = 1 \times 10^{20}$ electrons per m${}^2$ and
$M = 10,000$.  The extremely high density was chosen as this density not require significant
expansion of the bunch before the onset of relativistic effect.
Simulations were similar to those described in Fig. \ref{fig:1d ev}, and again the inset graphs
show the theoretic density at 2 much later times (0.5 and 1 ps).  
While the distributions at later times appear delta-like, they do have at least the same width
as seen at earlier time --- however, this width is much smaller than the scale resulting in
the delta-like behavior at later times.  Notice that in the main plot
the scales are consistent among graphs.  
Also notice that the extraction fields have little
effect on where the front of the bunches are after 1 ps but have dramatic effect on the bunch 
distribution ---
this is partially an artifact of the high density of the initial distribution that results in the front
of the distribution relatively quickly becoming relativistic with or without an extraction field.
In addition to the shape of the asymptotic density, the extraction field determines to what extent
the initial variance about the mean-field theory prediction is lost 
(an effect explained in the text.)}
\end{figure*} 

\begin{figure*}[tbh]
  \begin{center}
  \subfloat[$E_a = 0.5 ~ E_{T1}$]{\includegraphics[width=0.45\textwidth]{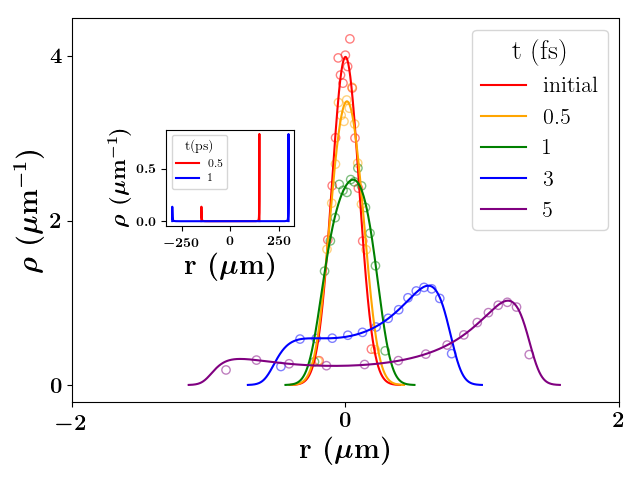}}
  \subfloat[$E_a = E_{T1}$]{\includegraphics[width=0.45\textwidth]{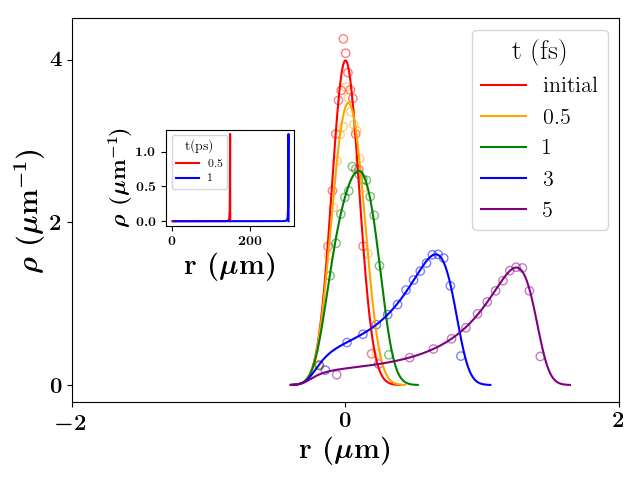}}\\
  \subfloat[$E_a = 2 ~ E_{T1}$]{\includegraphics[width=0.45\textwidth]{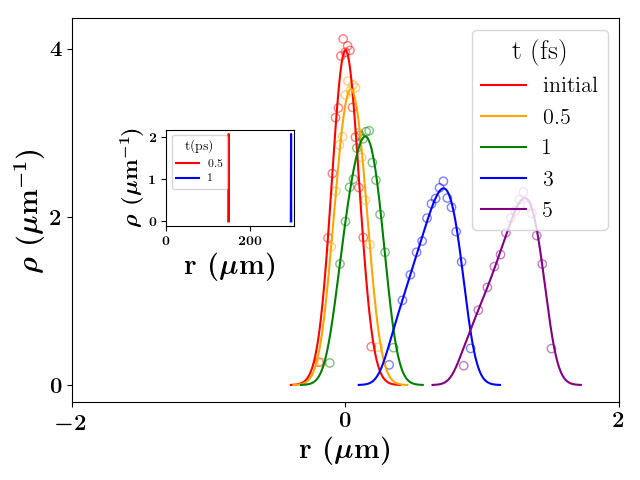}}
  \subfloat[$E_a = 10 ~ E_{T1}$]{\includegraphics[width=0.45\textwidth]{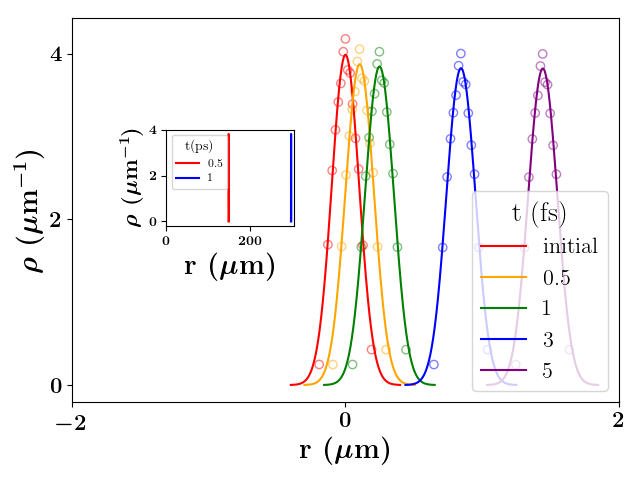}}
  \end{center}

\caption{\label{fig:Gaussian 1d ev} Theoretical predictions (solid line) and $M$-shell simulations (hollow dots)
of the planar symmetric density at 4 different times with different applied extraction fields.  
Parameters and simulations are analogous to those described in Fig. \ref{fig:uniform 1d ev} excepting
the initial Gaussian distribution with $\sigma_r = 0.1 ~\mu$m.}
\end{figure*}

The effect of the extraction field on the time-dependent density evolution can be seen in Figs.
(\ref{fig:uniform 1d ev}) and (\ref{fig:Gaussian 1d ev}), which show the evolution
of initially uniform and Gaussian distributions, respectively, in the presence of various extraction fields.
First, notice that the inclusion of a non-zero $E_a$ breaks the symmetry of the left and right pulses and 
that we can see that the double pulses are replaced by a single pulse as the applied field crosses
the virtual cathode limit, $E_a = E_{T1}$.  As $E_a$ is increased beyond
$E_{T1}$, all Lagrangian particles eventually become relativistic, and the density ``lifts" away from the axis.
Eventually (not shown), the extraction field should be strong enough that no appreciable expansion
occurs and the initial distribution is simply displaced at the speed of light; this can be shown to occur
when $E_a >>E_{T1}$.

Also as can be seen in Figs. (\ref{fig:uniform 1d ev}) and (\ref{fig:Gaussian 1d ev}), the initial
stochastic variation in the density is 
lost for simulations of sufficiently low extraction field but is retained for $E_a \ge 10 E_{T1}$.
This is due to the same effect discussed in the planar model without an electric field; however,
the relativistic time scale needs to be adjusted, namely 
\begin{align}
  \tau_{rel} = \frac{l_{r01}}{c}\left|P_{01} + \frac{E_a}{E_{T1}}\right|^{-1}\label{eq:1D timescale acc}
\end{align}  
When
$\tau_{rel} >> \tau_{exp}$, we again have the case where the expansion dynamics dominate and 
the inter-particle spacings essentially are equivalent to the inter-particle spacings determined by theory.
However, once $\tau_{rel} << \tau_{exp}$, the inter-particle spaces do not expand sufficiently to
overcome the initial stochastics and the variance is preserved.  The new wrinkle is that $\tau_{rel}$
can be reduced by simply increasing the extraction field.  Therefore, we do in fact see a distribution
evolve that retains the initial variance, i.e. $E_a = 10 E_{T1}$ in Fig. \ref{fig:uniform 1d ev}, as for that
simulation $\tau_{rel} << \tau_{exp}$.

Moreover, the influence of the extraction field is important in 
the highly relativistic regime not only for influencing the time scale but also influencing the 
asymptotic distribution.  Specifically, the effect of the extraction field in the 1D model is 
apparently not to accelerate the front of the distribution, i.e. all simulation had the front of the bunch
traveling near the speed of light, but instead to shape the eventual distribution
as can be seen in Figs. (\ref{fig:uniform 1d ev}) and (\ref{fig:Gaussian 1d ev}).  
 The asymptotic densities for the
initially uniform and Gaussian distributions and for various 
extraction fields can be seen in Fig. \ref{fig:asymptotes} where we have used Eq. (\ref{eq:rel asym}).
Specifically, every point besides
the point corresponding to $P_{01} + \frac{E_a}{E_{T1}} = 0$ will eventually have 
$ |P_{01} + \frac{E_a}{E_{T1}}| \frac{ct}{l_{r01}} >> 1$ and
thus the density corresponding to such points will eventually become a constant.  However, while
$|P_{01} + \frac{E_a}{E_{T1}}| \frac{ct}{l_{r01}}$ is not much larger than $1$, 
the density of the point will decrease toward
the eventual constant value.    

\section{Application of the 1D distribution}\label{sec:applied}
In this section, we demonstrate one use of the spatial distributions; specifically, we calculate the width 
evolution of UEM-relevant planar-symmetric distributions as a function of time.  
Specifically, we define the rms width of the distribution
as
\begin{align}
  \sigma_z &= \sqrt{<z^2> - <z>^2}
\end{align}
Theoretically $<a> = \int_{-\infty}^{\infty} a \rho_1 dz =  \int_{-\infty}^{\infty} a(z_0) \rho_{01} dz_0$, and in simulation $<a> = \frac{1}{N} \sum_{i=1}^N a_i$ where $N$ is
the number of particles in the simulation and $a_i$ is the value of $a$ for the $i^{th}$ particle.  

\begin{figure*}
  \begin{center}
 \subfloat[$10^6 ~e$]{\includegraphics[width=0.45\textwidth]{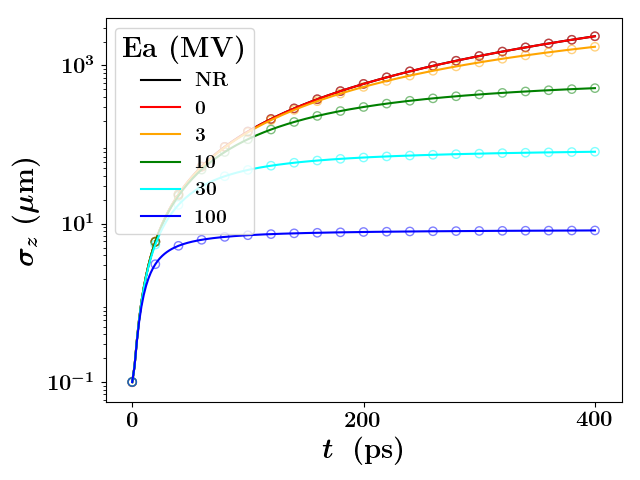}}
  \subfloat[$10^8 ~e$]{\includegraphics[width=0.45\textwidth]{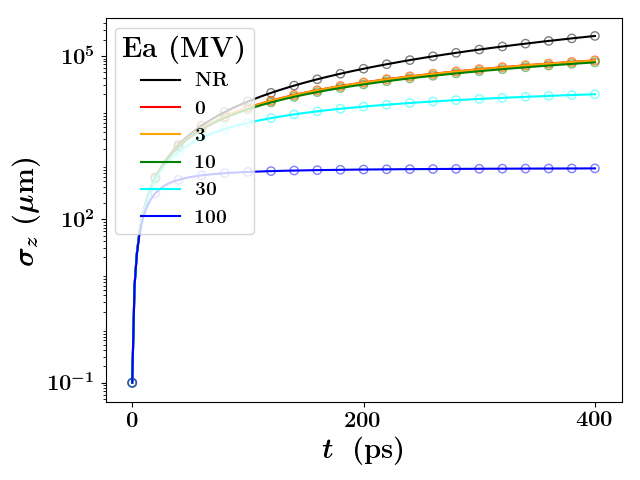}}
  \end{center}
  
\caption{\label{fig:width ev} Theoretical predictions (solid line) and $M$-shell simulations (hollow dots)
of the width of an evolving planar-symmetric initially Gaussian distribution 
for two experiment relevant regimes.  Both cases assumed a geometry of transverse width of
$100~\mu$m and longitudinal width of $0.1 ~\mu$m, and they differed in the number of electrons
in the bunch: (a) $10^6~e$ corresponding to an $E_{T1} \approx 0.3 MV$ and (b) $10^8~e$
corresponding to an $E_{T1} \approx 3 MV$.  Notice that the y-axis is logarithmic, and the scales
are different between the two plots. Also notice that relativistic effects of spreading of the bunch in 
the absence of an accelerating field within this time scale are not significant for the $10^6$ case
but noticeable for the $10^8$ case.  Furthermore, the accelerating field effect
is noticeable within this time scale when $E_a = 10 MV$ for $10^6$ and $E_a = 30 MV$ for $10^8$;
however, notice that in the $10^8$ case, the longitudinal beam width is larger than the transverse width
and that the $1D$ model is probably no longer a valid approximation.}
\end{figure*}

We consider a Gaussian bunch with transverse radius of $100 ~ \mu$m and longitudinal width of 
$\sigma_r = 0.1 ~ \mu$m, and we consider both $N=10^6$, relevant for diffraction studies, and 
$N=10^8$, relevant for imaging studies.
We treat the longitudinal expansion with the planar model both using the non-relativistic distribution,
Eq. (\ref{eq:1D ev non}), as well as the relativistic version, Eqs. (\ref{eq:1D rho}) 
and (\ref{eq:rel transform}). 
We calculate the theoretical expectation numerically for initially Gaussian
distributed planar-symmetric distributions for various values of $E_a$ as well as the non-relativistic
width prediction and compare the results to
the standard deviation of $M=10^4$-shell simulations with the same parameters.  
Results of this treatment may be seen in Fig. \ref{fig:width ev}.  

As can be seen in Fig. \ref{fig:width ev},
the theory and simulations result in the same width evolution.
It is worth noting that $E_{T1} \approx 0.3 MV/m$ for $N=10^6$ and $E_{T1} \approx 30 MV/m$ for $N=10^8$.
As can be seen in the figures, the width growth does not vary much from the unaccelerated case until 
an extraction field is increased beyond $\approx 10 ~ E_{T1}$, that is the expansion dynamics will
dominate the width determination until we are far beyond the ``total" field within this 400 ps time frame.  
Also apparent is that,
within this time frame, the dynamics of the un-accelerated bunch does not differ from the 
non-relativistic model;  on the other hand, the higher density dynamics do differ suggesting that
relativistic expansion occurs in the $N=10^8$ planar model.  Of course, as this bunch expands its 
longitudinal length will quickly become larger than the transverse width suggesting higher-dimensional
dynamics will become important.  Specifically after the transition to higher-dimensional dynamics, the planar 
model overestimates the longitudinal width and underestimates the transverse width.  
However, if a sufficient extraction field is obtained,
i.e. around $100 E_{T1}$, the asymptotic 
longitudinal width is of sufficiently small size to result in planar dynamics for the bunch meaning that
the bunch width can be modeled with the planar model in that regime.  

\section{Discussion and Conclusions}\label{sec:discussion}
In this work, we extended our previous density evolution analysis into the relativistic regime.  
Specifically, we showed that the uniform distribution in any dimension develops density shocks as the outer
portion of the distribution becomes relativistic; we also found expressions for such peaks in other
distributions which occur in competition with the non-uniform Coulomb mechanism that leads to peaks 
in such distributions\cite{Zerbe:2018_coulomb_dynamics}.  We showed that the analytic results accurately predicted 
1D-like $M$-shell simulation results under all 
symmetries and PIC results  under cylindrical and spherical symmetries.
The PIC simulations conducted here were completed
using an EM solver with an initial ES solve used to initialize the fields.  As these simulations
agree with the theory that is essentially based on electrostatics,  it is apparent
that EM effects beyond electrostatics are not significantly affecting the density evolution 
for the problems examined.

We emphasize that the mechanism for the relativistic shock development 
is distinct from non-relativistic shock development seen in the Gaussian 
distribution\cite{Zerbe:2018_coulomb_dynamics}.  Previously,
we demonstrated that shocks arise in non-planar non-uniform distribution evolutions
due to the initial distribution leading to non-linear Lagrangian particle velocities that lead
to inner Lagrangian particles catching up to outer Lagrangian particles.   On the other hand,
shock in a relativistic bunch are caused by the 
``shrinking" of one-dimension of the density as the Lagrangian particles approach the luminal speed limit.
This can be seen by considering the energy of continuum particles within the distribution.
Specifically, as the particles expand, their kinetic energies increase according to Eq. (\ref{eq:KE gen}).  In
the planar and cylindrically symmetric models, this increase is linear and logarithmic in their position
(and eventually time),
respectively, as can be
seen by Eqs. (\ref{eq:U1}) and (\ref{eq:U2}).  This leads to all particles in a neighborhood approaching the 
speed of light resulting in the ``freezing'' of the expansion along the expanding dimension; that
is, all planar symmetric distributions eventually asymptote to a time independent 
density while all cylindrically
symmetric distributions eventually expand ``uniform-like" but with one dimension less
than that being considered, i.e. $\rho_2 \to \frac{r_0}{r} A \rho_{02}$ where $A$ is some parameter
determined from the initial conditions.  On the other hand,
in the spherically symmetric case, the kinetic energy is bounded 
by $\frac{qQ_{tot} P_{03}}{4\pi \epsilon_0 r_0}$, which is finite.  
As this kinetic energy is dependent on $r_0$ through $\frac{P_{03}(r_0)}{r_0}$,
the asymptotic velocity of neighboring continuum particles differs.  This is why 
the density at long times for highly relativistic portions of the distribution 
drops in a uniform-like manner, i.e. $\rho_3 \to \frac{r_0^3}{r^3}A \rho_{03}$
with $A = \frac{1 + 3 \zeta^2 + 2 \zeta^4}{1 + \frac{3}{2} D_{03}}$ (see Eq. (\ref{eq:intro rhox})).  
While both the planar 
and cylindrically symmetric cases lost a power to the uniform-like evolution, i.e.
planar cases evolve like $\frac{1}{r^0}$ and cylindrically symmetric cases evolve
like $\frac{1}{r^1}$, the spherically symmetric case's uniform-like evolution retains $\frac{1}{r^3}$.  
This difference arises as the particles in the spherically symmetric case have finite potential energy 
and thus asymptote towards a velocity that is a little less than the speed of light.
Now neighboring Lagrangian particles can have very small differences in their asymptotic velocity
leading to the expansion in the radial direction being slower than what is seen non-relativistically; 
specifically, this
is what leads to $A > 1$.  However, there does remain a non-vanishing small velocity difference
meaning that the distribution continues to expand in all dimensions.  
Nonetheless, these effects lead to density peaks forming towards the edges of the distribution
in all cases;
regardless, we find it interesting that the behavior in each dimension is qualitatively unique.

Furthermore, we demonstrated that if the distribution is given enough time to expand, the 
stochastic effects in the initial distribution are overwhelmed
by the space charge effects.  This means that under such conditions, repeated instances of similar bunches
should look more or less the same.  However, if the bunch is quickly accelerated into the relativistic 
regime, the initial stochastic fluctuations are preserved.

While these results are somewhat surprising, we do need to emphasize that these conclusion are
based on analyzing the symmetric models 
at long times and that some of the physical assumptions inherent in the models 
should be violated at some point.  The three assumptions for these 
model are (1.) the temperature is small compared to the kinetic energy delivered to the particle
due to Coulomb
interaction, (2.) the distributions remain laminar, and (3.) the symmetry under consideration represents 
the physical situation.
While these assumptions all break down to some extent at some point, 
we emphasize that in the simulations we
have conducted that the model almost exactly matches with the PIC simulations.

Note that even as the distribution becomes more and more diffuse,
if a region of the distribution had much less heat than the kinetic energy delivered to it by space-charge
effects, we'd expect the particles' trajectories to 
not be drastically altered from the trajectory determined by the space-charge effects alone --- as long as
the potential energy is quickly converted to kinetic energy. 
Nonetheless, there is always 
a portion of the center of the distribution that does not meet this assumption.  In the planar
and cylindrically symmetric cases where the kinetic energy is unbounded,
this portion of the distribution is always shrinking; on the other hand, 
in the spherically symmetric case, there is
a portion of the distribution that will never become space-charge dominated
as the kinetic energy transferred to the particles in this region will never overcome the
energy associated with the initial temperature.   In other
words, in real world situations, the center of the
distribution is emittance dominated regardless of the fact that farther out in the distribution the 
particles may be space-charge dominated.  

 The second assumption of laminar behavior is surprisingly robust.  Obviously, having a higher temperature
 should lead to issues with this assumption, but for the cases we've examined
 within the temperature range where space-charge dominated fluid is present, this does not 
 seem to be much
 of an issue --- at least early on.  
 The biggest success of the laminar fluid assumption is the planar symmetry case
 as the acceleration of 
 successive sheets is monotonically increasing making laminar fluid assumption violating events impossible
 unless the initial velocity distribution is correctly tuned.   
 The real issue with this assumption, though, is with the non-uniform bunches under
 cylindrical and spherical symmetries.  As we have
 discussed in our previous paper, crossover events that violate the laminar fluid assumption occur 
 when $r' = 0$.  The density shock that ends up forming in the evolution of a non-uniform bunch
 can be thought of as occurring in region(s) of substantial initial density that have $r' \to 0$ relatively
 quickly.  In other words, successive cylindrical or spherical shells begin to bunch up as they expand
 resulting in a relatively higher density in those regions.  
 In the non-relativistic case, these shells eventually cross-over resulting in
 a violation of the laminar fluid assumption (although the model still predicts the density evolution fairly well 
 even past such events).  However, considering relativity, if the initial distribution is of sufficient density, 
 the expansion may be able to ``freeze" before this point. --- at least in the cylindrical case.  
 On the other hand, relativity does not
 help the spherically symmetric case as complete freezing never occurs.  Instead, as can be seen by
 analyzing Eq. (\ref{eq:limit sph}), for the region of the distribution where $D_{03} \approx -\frac{2}{3}$ all
 of the terms may be relevant.  As the term in front of the $\tanh^{-1}$ function is negative in this region 
 whereas the rest of the expression is positive, 
 there should be some value of $r_0$ that leads to $r' = 0$.  This tells us that at some time
 we should expect the laminar fluid assumption to be violated by a spherically symmetric distribution.  
 Of course for truly uniform distributions, $D_{03} = 0$ everywhere and
 this crossover does not happen,
 but for any realistic distribution, all values of $D_{03} < 0$ are present and crossover should
 occur.  Specifically, roughly 20\% of the Gaussian distribution has $D_{03}(r_0) < -\frac{2}{3}$ and 
 this crossover in this region apparently does not drastically change the evolution of the distribution for 
 the times
 we've considered in this paper although further examination of the behavior of the model in this
 region is warranted.
 Regardless, before the time of crossovers, we are confident that the dynamics of the
 distribution are captured by the expressions we have presented here.
 
 In the
 UEM community, the planar symmetric model is applied to a bunch that is thin along one
 axis with much larger widths along the other dimensions; we denote this as $L_0 << R_0$ where
 $L_0$ represents the initial width of the thin dimension and $R_0$ the initial 
 widths of the other two (equivalent)
 dimensions.  If planar symmetric dynamics are present, at some time $L \approx R$, and the 
 planar symmetric model should no longer apply instead requiring a higher dimensional
 description.  The time scale for the expansion of the bunch is
 $\tau_{exp} \approx \frac{1}{\omega_{01}}$; on the other hand, the time scale 
described by Eq. (\ref{eq:1D timescale acc}) indicates the time at which we would expect the edges of the 
distribution to have energy equivalent to the rest energy of the particle.  As we assume
$L_0 << R_0$, we'd expect that if these two timescales are of the same order or the relativistic
timescale is shorter than we'd expect
relativistic effects described by the models presented here to occur.  This occurs when $l_{r01} \le L_0$.
 Likewise,
 for the cylindrical case in fields like accelerator physics, 
 it is generally assumed $R_0 <<  L_0$; which again breaks down when
 $L \approx R$.  Nonetheless, we again expect the cylindrically symmetric dynamics described
 here to be apparent if $l_{r02} \le R_0$.
 
 It is straightforward to add an extraction field to the laminar theory.
For $10^8$ electrons in a uniform bunch of
radius $100 ~ \mu$m, $E_{T1} \approx 30 MV/m$; thus an acceleration field of 
$100 MV$, which is the upper limit of the UEM community used at 
the Stanford Linear 
Accelerator\cite{Musumeci:2010_single_shot,Weathersby:2015_slac,Murooka:2011_TED}, 
is only about $3.5 \times$ this quantity.  As we saw in Figs. (\ref{fig:uniform 1d ev}) and 
(\ref{fig:Gaussian 1d ev}),
in this range we would still expect space-charge effects that enact 
substantial expansion and distortion of the initial 
distribution.  On the other hand, table top UEM devices typically have extraction fields up
to $5 MV/m$\cite{Srinivasan:2003_UED,Ruan:2009_nanocrystallography,van_Oudheusden:2010_rf_compression_experiment,Sciaini:2011_review} , 
which is only slightly more than the total internal field of $10^7$ electrons in a pancake with
a radius of $100 ~\mu$m,
$E_{T1} \approx 3 MV/m$ and is far below $E_{T1}$ for $10^8$ electrons.  
Thus $10^8$ electron bunches are beyond the capability of such table top devices, and 
$10^7$ electron bunches should expand immensely within the extraction
field making them very difficult to work with.  

Notice that in previous treatments of the evolving 
density\cite{Reed:2006_short_pulse_theory,Zerbe:2018_coulomb_dynamics}, 
the extraction field was left
out of the analysis.  This
is accurate as the density evolution in
the non-relativistic limit, Eq. (\ref{eq:1D ev non}) is independent of the effective field.  However,
this is not true in the general case as relativistic effects make the electric field couple to the dynamics.
This leads to an interesting opportunity to control the density through this coupling 
effect.
Specifically, in 1D, the density freezing leads to the concept of asymptotic density, which is a density that no 
longer evolves in time.  We showed that this asymptotic density can be manipulated through 
the inclusion of an extraction field, $E_a$.  Specifically, the initial density is essentially the asymptotic
density when $E_a >> \frac{\Sigma_{tot}}{\epsilon_0}$; however, when the extraction field is not 
sufficiently large, the asymptotic density can be significant different from the initial density.  
This suggests that
if we are accelerating a bunch well into the relativistic regime, 
we may need to consider this asymptotic density
when determining optimal criteria.  Namely, in the relativistic regime, an initially uniform distribution
should no longer be the distribution with the smallest emittance as relativistic considerations
introduce non-linearities in the phase space that may be absent from correctly chosen
initial distributions.  We
will develop this idea further in future work.

\appendix
\section{Cylindrical symmetric density evolution in the highly relativistic regime}\label{ap:Cyl rel}

Assuming $2 {\tilde{\beta}}^2 >> 1$ and analyzing all but $\mathcal{F}$ and $\mathcal{F}_\partial$, 
Eq. (\ref{eq:cyl drdr0}) becomes
\begin{align}
  r' \approx \frac{r}{r_0} \left( 1 + \left(\frac{\rho_0}{{\bar{\rho}}_0}-1\right) \frac{1}{{\tilde{\beta}}}\mathcal{F} - \frac{\rho_0}{{\bar{\rho}}_0} \mathcal{F}_\partial\right)\label{eq:cyl drdr0 approx}
\end{align}
However,
\begin{align}
  \mathcal{F} &\approx \frac{r_0}{r} \int_{0}^{\sqrt{\ln\left(\frac{r}{r_0}\right)}} 2 {\tilde{\beta}} y e^{y^2} dy\nonumber\\
                      &=  {\tilde{\beta}}\left(1 - \frac{r_0}{r}\right)
\end{align}
and
\begin{align}
  \mathcal{F}_\partial &\approx \frac{r_0}{r} \int_{0}^{\sqrt{\ln\left(\frac{r}{r_0}\right)}} 2 y e^{y^2} dy\nonumber\\
                      &= 1 - \frac{r_0}{r}
\end{align}
Placing these approximations back into Eq. (\ref{eq:cyl drdr0 approx}) results in $r' \approx 1$.

\section{Long-time limit}\label{ap:long time}
Consider spherical symmetry.  Notice that $\frac{r_0}{r} r'$ is a function of $\frac{r_0}{r}$, so in the 
limit $\lim_{\frac{r}{r_0}\to\infty}$, these terms go to zero.  As a result, $x\to 1$ in the expression for $r'$,
so 
\begin{align}
  lim_{\frac{r}{r_0} \to \infty}r' &= \frac{r}{r_0}\left(1 + \frac{p_1(1)}{b_1^2 (g_2(1))^2} \right.\nonumber\\
              &\quad\quad \left.+ \frac{r_0}{r} \frac{p_2(1)}{b_1^2 (g_2(1))^2} T\left(\sqrt{1 - \frac{r_0}{r}}\right)\right)\nonumber\\
              &= \frac{r}{r_0}\left(\frac{(1 + \frac{3}{2} D )(1 +  {\tilde{\beta}}^2)}{(1 + {\tilde{\beta}}^2)^2(1 + 2 {\tilde{\beta}}^2)}\right) \nonumber\\
              &\quad\quad + \frac{(3 {\tilde{\beta}}^2 + 6 {\tilde{\beta}}^2 D + \frac{3}{2} D)\tanh^{-1}\sqrt{1 - \frac{r_0}{r}}}{(1 + {\tilde{\beta}}^2)^2(1 + 2 {\tilde{\beta}}^2)}\nonumber\\
              &\to \frac{r}{r_0}\left(\frac{1 + \frac{3}{2} D }{1 +  3 {\tilde{\beta}}^2 + 2 {\tilde{\beta}}^4 }\right)\label{eq:limit sph}
\end{align} 
where ${\tilde{\beta}} =  \frac{r_0{\bar{\omega}}_{03}}{\sqrt{6}c}$
and where the second term is lost since inverse hyperbolic tangent goes to infinity logarithmically 
which is slower than $\frac{r}{r_0}$.

\section{High density limit}\label{ap:high density}
At high densities, the edges of the planar symmetric distribution do not significantly evolve, 
and therefore the distribution is essentially preserved in this region.  This occurs when 
$\frac{2 \omega_{01} L_0}{c} >> 1$.  We now extend this to the other symmetries.

\subsection{Cylindrical symmetry}\label{ap:Cyl rel}

Assuming $\tilde{\beta} >> 1$, where ${\tilde{\beta}} =  \frac{r_0{\bar{\omega}}_{02}}{2c}$, 
and analyzing all but $\mathcal{F}$ and $\mathcal{F}_\partial$, 
Eq. (r') becomes
\begin{align}
  r' \approx \frac{r}{r_0} \left( 1 + \left(\frac{\rho_0}{{\bar{\rho}}_0}-1\right) \frac{1}{{\tilde{\beta}}}\mathcal{F} - \frac{\rho_0}{{\bar{\rho}}_0} \mathcal{F}_\partial\right)\label{eq:cyl drdr0 approx}
\end{align}
However,
\begin{align}
  \mathcal{F} &\approx \frac{r_0}{r} \int_{0}^{\sqrt{\ln\left(\frac{r}{r_0}\right)}} 2 {\tilde{\beta}} y e^{y^2} dy\nonumber\\
                      &=  {\tilde{\beta}}\left(1 - \frac{r_0}{r}\right)
\end{align}
and
\begin{align}
  \mathcal{F}_\partial &\approx \frac{r_0}{r} \int_{0}^{\sqrt{\ln\left(\frac{r}{r_0}\right)}} 2 y e^{y^2} dy\nonumber\\
                      &= 1 - \frac{r_0}{r}
\end{align}
Placing these approximations back into Eq. (\ref{eq:cyl drdr0 approx}) results in $r' \approx 1$.  That is,
$r'$ cancels out the factor $\frac{r}{r_0}$ term.

\subsection{Spherical symmetry}
Assuming $\tilde{\beta} >> 1$, where ${\tilde{\beta}} =  \frac{r_0{\bar{\omega}}_{03}}{\sqrt{6}c}$,
we see that Eq. (\ref{eq:limit sph}) is approximately $0$; therefore we need to return to the full expression
and expand in terms of $\frac{r}{r_0}$.  We find only keeping the highest order of ${\tilde{\beta}}$
\begin{align}
  r' &\approx \frac{r}{r_0}\left(1 + \frac{-2 {\tilde{\beta}^2}\left(1 - \frac{r_0}{r}\right)^2}{2 {\tilde{\beta}^6}\left(1 - \frac{r_0}{r}\right)}\right)\nonumber\\
     &= \frac{r}{r_0} \left(1 - \left(1 - \frac{r_0}{r}\right)\right)\nonumber\\
     &= 1
\end{align}
So like the planar and cylindrical symmetric cases, 
super-highly relativistic densities result in essentially the loss of one dimension
during expansion.

\bibliographystyle{apsrev4-1}
\bibliography{../coulomb_dynamics_APS/CoulombDynamics.bib}

\end{document}